\documentclass[prb,amsfonts,amssymb,floats,twocolumn,superscriptaddress,aps]{revtex4-2}
\usepackage[T1]{fontenc}
\usepackage{amsmath}
\usepackage{float}
\usepackage[caption = false]{subfig}
\usepackage{color}
\usepackage{braket}
\usepackage{bm}
\usepackage{dsfont}
\usepackage{adjustbox}
\usepackage{graphicx}
\usepackage{enumitem}
\usepackage{bbm}
\usepackage{tikz}
\usetikzlibrary{patterns}

\usepackage[colorlinks=true]{hyperref}
\hypersetup{linkcolor=blue,citecolor=blue,urlcolor=blue}

\renewcommand{\Re}{\operatorname{Re}}
\renewcommand{\Im}{\operatorname{Im}}

\newcommand{\iu}{\mathrm{i}} 
\newcommand{\eu}{\mathrm{e}} 
\newcommand{\du}{\mathrm{d}} 
\newcommand{\hc}{\mathrm{h.c.}} 
\newcommand*\Diff{\mathop{}\!\mathcal{D}}

\newcommand{\ver}{{\bf{r}}}
\newcommand{\vhat}[1]{\bm{\hat{#1}}}
\newcommand{\uc}{\upsilon_{\rm u.c.}}
\newcommand{\ve}[1]{{\boldsymbol #1}}
\newcommand{\Ef}{\vec{\mathcal{E}}}

\newcommand{\Uone}{\mathrm{U(1)}}

\newcommand{\SO}{\mathrm{SO}}

\newcommand{\nips}[0]{\mathrm{NiPS}_3}
\newcommand{\non}{\nonumber}

\DeclareMathOperator{\tr}{tr}

\DeclareMathOperator{\re}{Re}
\DeclareMathOperator{\im}{Im}

\graphicspath{{./}{figs/}}

\begin{document}

\title{Ultrafast optical excitation of magnetic dynamics in van der Waals magnets: Coherent magnons and BKT dynamics in $\nips{}$}



\author{Urban F. P. Seifert}
\thanks{These two authors contributed equally.}
\affiliation{Univ Lyon, ENS de Lyon, Univ Claude Bernard, CNRS, Laboratoire de Physique, 69342 Lyon, France}
\affiliation{Kavli Institute for Theoretical Physics, University of California, Santa Barbara, CA 93106, USA}

\author{Mengxing Ye}
\thanks{These two authors contributed equally.}
\affiliation{Kavli Institute for Theoretical Physics, University of California, Santa Barbara, CA 93106, USA}

\author{Leon Balents}
\affiliation{Kavli Institute for Theoretical Physics, University of California, Santa Barbara, CA 93106, USA}
\affiliation{Canadian Institute for Advanced Research, Toronto, Ontario, Canada}

\date{\today}

\begin{abstract} 
Optical pump-probe experiments carried out in the time domain reveal both the intrinsic low energy dynamics and its connections to higher energy excitations in correlated electron systems.  
In this work, we propose two microscopic mechanisms for the optical generation of coherent magnetic modes in van der Waals magnets, and derive the corresponding effective light-spin interactions: either through pumping atomic orbital excitations resonantly or via a light-induced Floquet spin Hamiltonian, the ground state of the system is driven out of equilibrium.
The subsequent long-time relaxational dynamics can then be probed using, e.g.\ the magneto-optical Kerr effect or transient grating spectroscopy.
As an example, we apply our framework to $\nips{}$, which is magnetically ordered in the bulk, and is conjectured to realize the XY model in the monolayer limit.
Our theory makes explicit how the material's low-energy response depends sensitively on the microscopic details of the light-spin coupling as well as pump fluence, frequency and polarization.
For the case of bulk $\nips{}$, we find quantitative agreement with recent experiments by Afanasiev et al.\ in Ref.~\onlinecite{afanasiev20}.
We further propose pump-probe experiments for monolayer $\nips{}$ and detail how anomalous relaxational behaviour may reveal excitations of a (proximate) BKT phase in a proposed effective XY model.
\end{abstract}
\maketitle


\section{Introduction}
Ultrafast light-matter interaction provides a powerful means to probe and control quantum materials \cite{kiri10,baso17,TorreReview2021}.
By pumping the system to a non-equilibrium state, the probed relaxation dynamics may reveal intrinsic coherent excitations.
Recent advances in the field of ultrafast optics have enabled coherent control of quantum materials through non-thermal pathways, where the quantum coherence in the transition process must be taken into account.
While these non-thermal pathways enable more control by the pump polarization, frequency, fluence, etc., they also pose the theoretical challenge of identifying microscopic mechanisms suitable for pumping in a specific setting.
Several mechanisms have been explored both experimentally and theoretically to optically excite coherent magnetic excitations~\cite{fiebig04,tzsch17,seiba19}, e.g.\ through coupling with optically active phonons, or direct coupling with spin excitations.
Recent experiments also showed coherent magnetic excitations through pumping in the mid-infrared to near-infrared range, where the light couples to correlated electronic degrees of freedom or atomic excitations.

Moving a step forward towards a \emph{systematic} understanding of the microscopic pathways of optical generation of coherent magnetic excitations, in this study we explore two distinct microscopic mechanisms, one resonant and another non-resonant, through optically pumping atomic orbital excitations and Floquet engineering, respectively. 

Our studies are motivated by recent experiment by Afanasiev et al. in Ref.~\onlinecite{afanasiev20}, which point towards markedly distinct mechanisms of exciting coherent magnons in the same material.
Using a pump-probe setup, they find that applying linearly polarized pump beams to the bulk van der Waals magnet $\nips{}$ allows one to selectively excite two distinct modes of oscillations in the Faraday rotation of the probe beam as signature of the coupled dynamics of the Néel order parameter and magnetization.

$\nips{}$, which belongs to the family transition-metal thiophosphates has zigzag-antiferromagnetic Néel order in the ground state \cite{wild15,lan18,gu19}.
A single-ion easy-plane anisotropy causes the spins to lie within a plane, with a weaker easy-axis anisotropy due to monoclinic stacking of layers setting an in-plane ordering direction (for the definition of the spin Hamiltonian and modeling of the magnon dynamics we refer the reader to Sec.~\ref{sec:bulknips}).
The low-energy magnon modes of the system thus correspond to pseudo-Goldstone modes.
The lowest-energy mode is associated with rotations of the Néel vector against the in-plane easy-axis anisotropy, dubbed $f_1$ in Ref.~\onlinecite{afanasiev20}, while fluctuations of the out-of-plane component give rise to a higher frequency mode $f_2$. The former is found to be excited for pump beam energies in a narrow frequency range around $1.0\,\mathrm{eV}$, close to the energy of an atomic orbital transition.
However, strikingly, the out-of-plane mode is excited with an almost equal amplitude for a wide range of photon energies from $0.1\,\mathrm{eV}$ up to $0.9\, \mathrm{eV}$.
This remarkable difference in the dependence of the excited modes on the driving light's frequency suggests, in the spirit of the discussion above, that two distinct excitation mechanism are at play.

In this work, we show that the optical pumping of orbital atomic excitations is efficient to drive the resonant $f_1$ mode. Furthermore, we show that the transient Floquet spin Hamiltonian allows for additional anisotropic terms, which disturb the zigzag ground state ordering pattern.
This can explain the $f_2$ mode being excited in a transparent energy window.
We emphasize that spin-orbit coupling is crucial to optically excite magnetic excitations through electronic transitions. 

Both the microscopic mechanisms as well as the theoretical framework to describe the ultrafast excitation process discussed here can be readily generalized to pumping other types of magnetic excitations.
As a second example, we hence discuss pumping hydrodynamic modes in the Berezinskii-Kosterlitz-Thouless (BKT) phase and its proximate phases.
This may be realized in the monolayer limit of $\nips{}$, which was found to exhibit a drastic suppression of antiferromagnetic ordering compared with two-layer and bulk samples~\cite{kim19}.

\subsection{Framework} \label{sec:framework}

Our methodology in modeling the microscopic excitation mechanism of coherent magnetic excitations in ultrafast pump-probe setups consists in finding an effective microscopic Hamiltonian $\mathcal{H}_\mathrm{eff}$ which is active for the duration of the pump in second-order perturbation theory in the electric field of the driving light.
Time evolution according to this Hamiltonian takes the system out of equilibrium and thus sets the initial conditions for the subsequent relaxation dynamics according to the equilibrium low-energy equations of motion.

In the case of bulk $\nips{}$, the homogeneous low energy excitations consist of two sets of coupled harmonic oscillators
\begin{equation}
	\partial_t n_\alpha \sim \chi^{-1} m_\beta - h_{m,\beta}(t) \quad \partial_t m_\beta \sim - \kappa_{n_\alpha} n_\alpha + h_{n,\alpha}(t),
	\label{eq:nmEoM}
\end{equation}
for the components of the Néel vector $\bm n$ and the magnetization $\bm m$, with $\alpha = y(z)$ and $\beta = z(y)$ for the in-plane $f_1$ (out-of-plane $f_2$) mode when the N\'eel order is along $x$. 
The source terms $h_{m,\beta}$ and $h_{n,\alpha}$ correspond to effective uniform and staggered magnetic fields induced by the pump beam's electric field, and can be obtained by considering the action of $\mathcal{H}_\mathrm{eff}$ on the low-energy variables of the system.
We note that this approach presupposes that the time-evolution induced by the pump pulse is unitary, and thus that the concept of a effective (hermitian) Hamiltonian is well-defined.
Interestingly, this assumption is {\em not} satisfied for driving frequencies which are resonant with a (sharp) atomic transition, as we discuss in Sec.~\ref{sec:tdpt}.
In this case, the initial conditions for the low-energy variables set by the perturbing light can be evaluated directly in a microscopic picture using a generalized (non-unitary) time-evolution operator.  Eqs.~\eqref{eq:nmEoM} nevertheless still hold (with $h_{\cdot}=0$) after the pulse.

With an advanced understanding of the microscopic mechanisms governing the excitation of low-energy magnons in conventionally ordered magnetic systems at hand, it is of prime interest to turn to more exotic magnetic phases of matter.
Here, ultrafast optical methods may serve as an additional probe into the dynamics of more exotic excitations beyond the linear response regime.
To this end, we study pumping hydrodynamic modes in the BKT phase and its proximate phases.
The general phase diagram of a 2D magnet with easy-plane anisotropy and hexagonal symmetry consists of two phase transition at $T_{\rm BKT}$ and $T_{\rm c}$, where $T_{\rm BKT}>T_{\rm c}$~\cite{Jose1977}.
Below $T_{\rm c}$, the system orders into one of the six degenerate minima selected by the hexagonal anisotropy. Above $T_{\rm BKT}$, the system is disordered with exponential decaying correlations.
At any $T_{\rm c}<T<T_{\rm BKT}$, the system exhibits critical algebraic order.
Crucially, the pump-induced excitation and subsequent relaxation of spin-wave modes in the (quasi-)ordered phases and diffusive spin modes in the disordered phases can be readily modelled using the framework described above: The time evolution according to the effective Hamiltonian which is active during the pump, with the respective light-induced local anisotropies and exchange interactions between the $S=1$ local moments, then sets the initial conditions for the subsequent low energy dynamics of the in-plane phase of the Néel order parameter and the out-of-plane magnetization, which are conveniently modelled in terms of a \emph{dual} electromagnetic theory in order to account for the presence of free vortices and bound vortex-antivortex pairs \cite{ambe80}.

We emphasize that our framework can be readily applied to other systems with hexagonal symmetry and spin-orbit-coupling, with only minor modifications required to account for the possibly distinct crystalline symmetries. Furthermore, coherent magnetic excitations besides magnon and hydrodynamic modes may be excited within this framework, such as quadrupolar waves proposed in NiGa$_2$S$_4$~\cite{Stoudenmire2009}. 

The remainder of the paper is organized as follows.
In Sec. \ref{sec:atomic}, we discuss single-ion multiplets of a Ni$^{2+}$ ion in appropriate crystal fields and derive electric field-induced single-ion anisotropies.
The modification of (anisotropic) exchange interactions in the pump period is discussed in Sec.~\ref{sec:floquet}.
We then model the long-range magnetic order and the light-induced dynamics of coherent low-energy magnons in bulk $\nips{}$, using the results of the preceding Sections in Sec.~\ref{sec:bulknips}. In Sec.~\ref{sec:app_bkt}, we generalize the framework to pumping hydrodynamic modes in and proximate to the BKT phase.

\section{Pump-induced single-ion anisotropy}
\label{sec:atomic}

\begin{figure}
    \centering
    \includegraphics[width=.95\columnwidth]{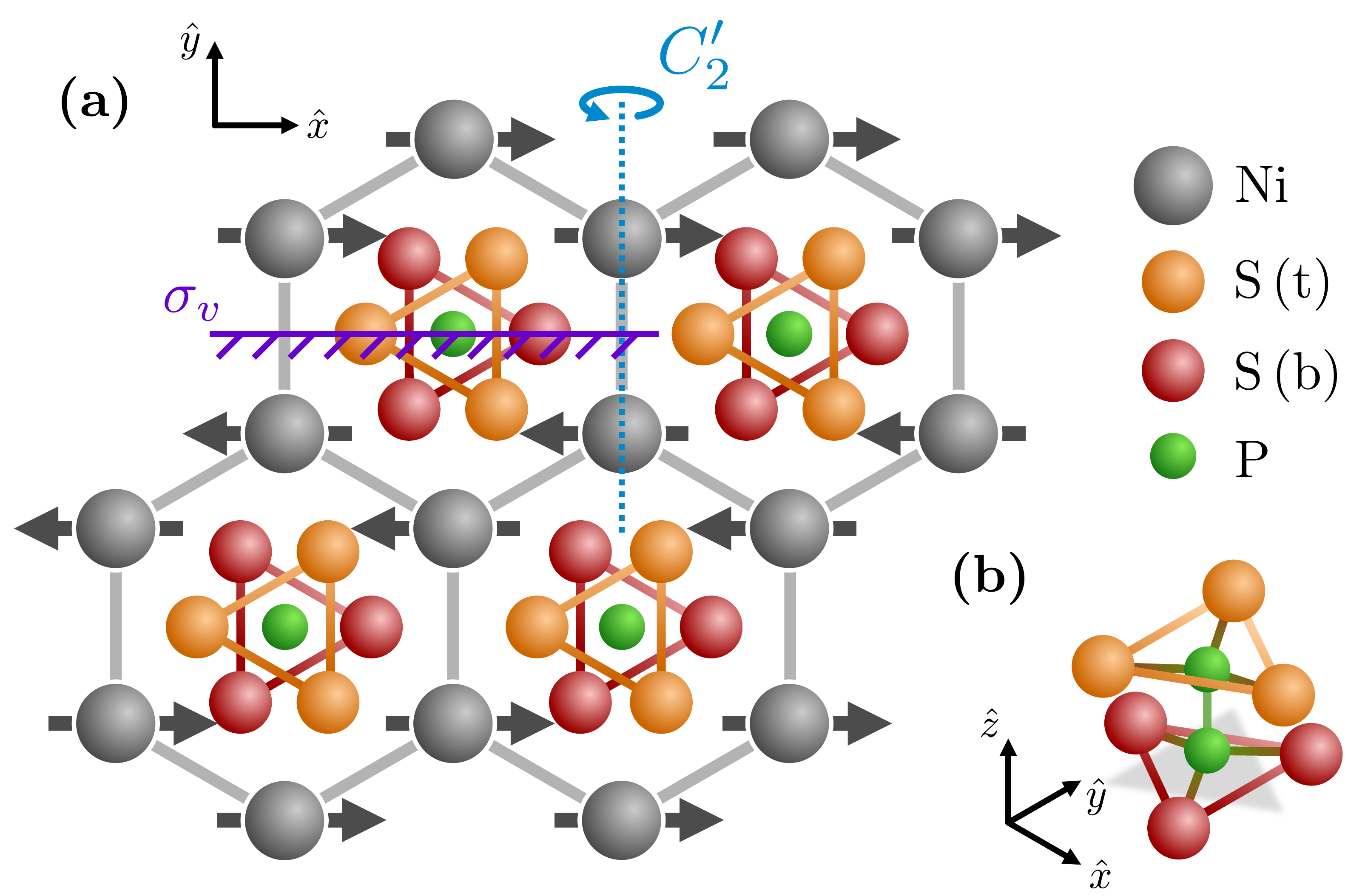}
    \caption{Schematic visualization of NiPS$_3$ crystal structure. (a) Projection in the $\hat{x}$-$\hat{y}$-plane, where S\,(t) and S\,(b) denote sulphur atoms on the top and bottom of a $(\mathrm{P}_2 \mathrm{S}_6)^{4-}$ cluster. The grey arrows indicate the zigzag Néel due to the $\mathrm{Ni}^{2+}$ local moments arranged on a honeycomb lattice. The axis for the $C_2'$ symmetry is shown marked light blue, and the vertical mirror plane $\sigma_v$ in purple. We also show the non-cubic crystal field environment (with broken inversion symmetry) for a Ni site. (b) Three-dimensional (isometric projection) illustration of a $(\mathrm{P}_2 \mathrm{S}_6)^{4-}$ cluster.}
    \label{fig:nips3_struct}
\end{figure}

Motivated by the observation in Ref.~\onlinecite{afanasiev20} that the low-energy in-plane mode $f_1$ in $\nips$ is excited in a rather narrow pump photon energy range which coincides with the energy of the $^3 A_{2g}\to {}^3 T_{2g}$ inter-orbital resonance of a Ni$^{2+}$ ion, we first focus on a single ion in a $d^8$ configuration in an octahedral crystal field environment (with a small trigonal distortion).

The laser's electric field induces virtual transitions out of the orbital singlet, spin triplet ground state to the first excited multiplet.
As the latter is split under spin-orbit coupling, we can use time-dependent second-order perturbation theory to derive an effective time evolution in the $S=1$ ground state sector which includes light-induced single-ion spin anisotropy terms.
Here we emphasize that spin-orbit coupling is a crucial ingredient as it allows the orbital resonance to couple to spin, which ultimately yields transitions between distinct spin states in the ground state manifold.
We further note that within our local approach of considering a single Ni$^{2+}$ ion with octahedral symmetry, all dipole matrix elements between the ${}^3 A_{2g}$ ground state multiplet and the first excited ${}^3 T_{2g}$ multiplet vanish by the dipole selection rule, as the electric-field induced dipole operator $r^\alpha \overset{\mathcal{I}}{\to} r^\alpha$ is odd under inversion $\mathcal{I}$, while $A_{2g}$ and $T_{2g}$ are even under $\mathcal{I}$, such that $\braket{A_{2g}|r^\alpha|T_{2g}} = -\braket{A_{2g}|r^\alpha|T_{2g}} \equiv 0$.
However, this an artefact of the single-ion picture, since in NiPS$_3$, the sites of the Ni$^{2+}$ are not inversion centers as the surrounding S atoms, stemming from the P$_2$S$_6$$^{4-}$ clusters, break the cubic symmetry, as visible from Fig.~\ref{fig:nips3_struct}. In this structure, the symmetry of the Ni sites is hence reduced to $D_3$. 
We therefore emulate the inversion-symmetry breaking by constructing a ground-state wavefunction for the $A_2$ orbital singlet which contains inversion-odd components due to the hybridisation induced by  ligand holes on the Ni$^{2+}$.
The induced anisotropy terms take the Néel vector out of equilibrium for the duration of the pump pulse and thus set initial conditions for the subsequent relaxation dynamics governed by the equilibrium equations of motion.

\subsection{$\mathrm{Ni}^{2+}$ single-ion multiplets}

To set the stage, the $\mathrm{Ni}^{2+}$ ion is in the local crystal field environment with $D_3$ symmetry, which is weakly broken from the cubic symmetry $O_h$ due to trigonal splitting. We note that under an octahedral crystal field the $^3 F$ ground-state multiplet of two unpaired $d$-electrons ($d^8 \simeq d^2$) is split as $^3 F \to {}^3 A_{2g} + {}^3 T_{1g} + {}^3 T_{2g}$, with ${}^3 A_{2g}$ constituting the $S=1$, orbital singlet ground-state manifold and ${}^3 T_{2g}$ the first excited multiplet.
To explicitly construct the corresponding wavefunctions from the single-particle $d$-orbital wavefunctions, it is convenient to choose the single-particle wavefunctions as eigenstates of the $C_3$ rotation operator about the $[111]$ axis in the cubic reference frame, where $\omega = \eu^{\iu 2 \pi /3}$ denotes the third root of unity.

The orbital component of the ground state is then given as $\ket{A_{2g}} = \frac{1}{\sqrt{2}} \left(\ket{e_{\omega^2}}_1 \ket{e_{\omega}}_2 - \ket{e_{\omega}}_1 \ket{e_{\omega^2}}_2 \right)$, and the orbital sector of the first excited ${}^3 T_{2g}$ multiplet is spanned by $\ket{T_{2g},1},\ket{T_{2g},\omega}$ and $\ket{T_{2g},\omega^2}$, with explicit expressions given in Eq.~\eqref{eq:T2g_c3}.
The energy of the first excited multiplet on top of the ground state is given by  $\varepsilon_{T_{2g}} - \varepsilon_{A_{2g}} = \Delta \equiv 10 \mathrm{Dq}$.

We now discuss the effect spin-orbit coupling $\mathcal{H}_\mathrm{SOC} = \lambda \vec L \cdot \vec S$ perturbatively.
As the ${}^3 A_{2g}$ multiplet is an orbital singlet, its threefold degeneracy will only be lifted at higher order perturbation theory through mixing with the excited levels.
Conversely, the degeneracy of the ${}^3 T_{2g}$ multiplet is lifted under spin-orbit coupling at first order in $\lambda$.
For our purposes, it is convenient to define a fictitious angular momentum $l=1$ acting on the $T_{2g}$ orbital triplet.
An explicit calculation reveals that
\begin{equation} \label{eq:L_projected}
	\mathcal{P}_{T_{2g}} \vec L \mathcal{P}_{T_{2g}} = \frac{1}{2} \vec l,
\end{equation}
where $\mathcal{P}_{T_{2g}} = \sum_{a = 1,\omega,\omega^2}\ket{T_{2g},a} \bra{T_{2g},a}$ denotes the projection operator into the orbital triplet.
At first order in $\lambda$, the degeneracy in the first excited multiplet is split with the effective Hamiltonian $\mathcal{H}^\mathrm{eff}_\mathrm{SOC}$.
Introducing the (fictitious) total angular momentum $\vec J_\mathrm{eff} = \vec S + \vec l$, the three sectors have quantum numbers $J_\mathrm{eff} = 0$, $J_\mathrm{eff} = 1$, $J_\mathrm{eff} = 2$ and energies $\varepsilon = -\lambda, -\lambda/2$ and $ \lambda/2$, respectively.

It is convenient to transform spatial coordinates to the trigonal reference frame with an orthogonal matrix $W$
\begin{equation} \label{eq:trafo-trigonal}
		\begin{pmatrix}x \\ y \\z \end{pmatrix} \mapsto \begin{pmatrix}\frac{2 z - x -y}{\sqrt{6}}\\ \frac{x-y}{\sqrt{2}}\\ \frac{x+y+z}{\sqrt{3}} \end{pmatrix} \equiv W\begin{pmatrix}x \\ y \\z \end{pmatrix} ,
\end{equation}
such that the quantization axes of spin and fictitious angular momentum operators coincide with the $\hat{z}$-axis, which is also the axis of the $C_3$ rotation operation.
Note that with this choice of reference frame, the splitting ${}^3 T_{2g} \to {}^3 A_{1} + {}^3E$ (for $\lambda =0$) under a trigonal distortion of the crystal field can be accounted for in degenerate perturbation theory through the effective Hamiltonian
\begin{equation}
	\mathcal{H}^\mathrm{eff}_\mathrm{tri} = \delta \left(l^z\right)^2.
\end{equation}
For both $\lambda, \delta \neq 0$ the ${}^3 T_{2g}$ multiplet is split into three doublets and three singlets which carry quantum numbers $S^z + l^z$.
However, for simplicity, we take $\delta = 0$ in the present discussion and note that a finite (but small) $\delta \neq 0$ could account for further level splitting and a fine structure of the experimentally observed resonance peaks.

\subsection{Time-dependent perturbation theory} \label{sec:tdpt}

We now employ time-dependent perturbation theory to study how the degeneracy of the $S=1$, orbital-singlet ground state sector spanned by $\ket{A_{2g},S^z = 0,\pm 1}$ is lifted by the perturbing electric field, which induces transitions to the $J_\mathrm{eff} = 0, 1, 2$ manifolds arising from spin-orbit splitting of the first excited multiplet.
To this end, we consider the perturbing Hamiltonian
\begin{equation}
	\mathcal{H}^\mathrm{E} = r^\alpha \mathcal{E}_\alpha(\omega) \eu^{\iu \omega t} + r^\alpha \mathcal{E}^\ast_\alpha(\omega) \eu^{-\iu \omega t}
\end{equation}
which yields the time-evolution operator in the interaction picture
\begin{equation}
	U_\mathrm{I}(t,t_0) = \mathcal{T} \eu^{-\iu \int_{t_0}^t \mathcal{H}^\mathrm{E}_{\rm I}(t') \du t'}.
\end{equation}
Expanding to second order and projecting to the $S=1$ subspace, we find the matrix elements of the effective time evolution operator within that subspace as
\begin{align}
    &\braket{a|U_\mathrm{eff}(t,t_0)|b} \approx \delta_{a b} \nonumber \\ &- \sum_m\int^t_{t_0} \du t' \int^{t'}_{t_0} \du t'' \braket{a|\mathcal{H}_\mathrm{I}^\mathrm{E}(t')|m}\braket{m|\mathcal{H}_\mathrm{I}^\mathrm{E}(t'')|b}, \label{eq:ueff-1}
\end{align}
where $\ket{a}\equiv \ket{S^z}$ denote the three states in $S=1$ ground-state sector and $\ket{m} = \ket{J_\mathrm{eff},J^z_\mathrm{eff}}$ are the states in the spin-orbit split first excited multiplet.
Focusing on a pump pulse of length $t_p$ which is switched on at $t_0 = 0$, the temporal integrations can be performed analytically, yielding
\begin{align} \label{eq:UI_analytical}
    \braket{a|U_\mathrm{eff}(t_p,0)| b} \approx \delta_{a b} + \mathcal{E}_\alpha \mathcal{E}^\ast_\beta \sum_m \frac{\braket{a|r^\alpha|m}\braket{m|r^\beta|b}}{\iu (\omega - \varepsilon_{m0})} \nonumber\\ \times \left[t_p - \frac{\eu^{\iu (\omega-\varepsilon_{m0})t_p}-1}{\iu (\omega- \varepsilon_{m0})} \right],
\end{align}
where we have kept only the dominant term for $\omega \approx \varepsilon_{m0} >0$ with $\varepsilon_{m0} = \varepsilon_m - \varepsilon_0$ denoting the energy difference between the ground-state and the first excited sector.

In the off-resonant limit, i.e. when $t_p(\omega - \varepsilon_{m0}) \gg 1$ for all $m$, the second term in the square brackets in \eqref{eq:UI_analytical} tends to 0, and the time-evolution operator can be written in terms of a pump-induced effective (hermitian) Hamiltonian, $U_\mathrm{I}(t_p,0) \approx \mathds{1} - \iu \mathcal{H}_\mathrm{eff} t_p$, as discussed by Pershan et al. in Refs.~\onlinecite{vdz65,pershan66}.

On the other hand, in the resonant limit $t_p(\omega-\varepsilon_{m0}) \ll 1$ it is seen that the time-evolution operator
\begin{equation}
    \braket{a|U_\mathrm{eff}(t_p,0)| b} \to \delta_{ab} - \mathcal{E}_\alpha \mathcal{E}_\beta^\ast  \sum_{m} \braket{a | r^\alpha |m} \braket{m| r^\beta | b} \frac{t_p^2}{2}
\end{equation}
becomes non-unitary, as the ground-state wavefunction acquires a finite overlap with the excited levels due to real transitions mediated by the perturbation, which is not captured within our  approach of projecting to the $S=1$ states.

Near a resonance, the notion of an effective pump-induced time-independent Hamiltonian thus no longer applies.
Instead, we will make use of the time-evolution operator projected to the low-energy subspace to directly compute the pump-induced initial conditions for relaxational low-energy dynamics as described in Sec.~\ref{sec:bulknips}.

In order to evaluate the time-evolution operator in the $S^z= \pm 1,0$-basis of the ${}^3 A_{2g}$ ground state, we first note that the equilibrium Hamiltonian $\mathcal{H}_0$ is diagonal following above considerations, and has elements
\begin{align}
	\mathcal{H}_0 \ket{A_{2g},S^z} &=0 \\
	\mathcal{H}_0 \ket{J_\mathrm{eff},J^z_\mathrm{eff}} &= \left(\Delta + \frac{\lambda }{4} J_\mathrm{eff}(J_\mathrm{eff}+1) - \lambda \right)  \ket{J_\mathrm{eff},J^z_\mathrm{eff}},
	\label{eq:H0m}
\end{align}
where $J_\mathrm{eff} = 0,1,2$.
Using the multiplet wavefunctions constructed in the previous section however immediately yields $U_\mathrm{I} \approx \mathds{1}$ as all dipole matrix elements between $d$-orbitals in the nominator of \eqref{eq:UI_analytical} vanish, in apparent contradiction to the experimentally observed orbital resonance.
To resolve this inconsistency, we note that recent experimental and first-principle numerical studies \cite{kim18,kang20} have recently shown that the $\mathrm{Ni}$ ground state contains -- in addition to the $d^8$ configuration -- a strong admixture of a $d^9 \underline{L}$ configuration, where $\underline{L}$ denotes a ligand $p$-orbital hole.
As the numerical modeling of a $\mathrm{Ni}$-ligand cluster in Ref.~\onlinecite{kim18} finds the hole-doped contribution to the ground state wave function $\ket{\psi}$ to be actually dominant $|\braket{d^9 \underline{L}| \psi}|^2 =0.60$ (compared to $|\braket{d^{10} \underline{L}^2| \psi}|^2 =0.15$ and $|\braket{d^8| \psi}|^2 =0.25$), we in the following consider the ground state configuration $e_g \underline{L}$ comprised of an $e_g$ and $\underline{L}$ hole.
We emphasize that in the case of octahedral symmetry the resulting ground-state wavefunction is found to be of (approximate) ${}^3 A_{2g}$ symmetry and thus -- given the oddness of the dipole operator appearing in the perturbing electric-field Hamiltonian $\mathcal{H}_\mathrm{I}^\mathrm{E}$ -- all dipole matrix elements for transitions to the first excited multiplet remain forbidden (however, trigonal distortions the $\underline{L}$ hole may induce an $e_u$ representation at the $\mathrm{Ni}^{2+}$ site).
Given that in $\nips{}$ the $\mathrm{Ni}$ are no longer centers of inversion of the $\mathrm{Ni}\mathrm{S}_6$ clusters, the $e_u$ orbitals due to the ligand hole may hybridize with the $e_g$ deriving from the $\mathrm{Ni}^{2+}$ $d$-levels to form an inversion-odd component in the full $A_2$ orbital singlet ground state given by
\begin{equation}
	\ket{A_{2}} = \frac{1}{2}\left(\ket{e_\omega}_1\ket{p_{\omega^2}}_2 + \ket{e_{\omega^2}}_1\ket{p_{\omega}}_2 + \left(1 \leftrightarrow 2\right) \right),
\end{equation}
where we drop all other contributions which will yield vanishing transition dipole matrix elements.

We are now in a position to evaluate \eqref{eq:UI_analytical} where we take $\ket{A_{2g},S^z} \to \ket{A_2}\ket{S^z}$.
We consider a pump beam at normal incidence to basal planes of $\nips{}$, implying (in the trigonal  reference frame) that $\mathcal{E}_z \equiv 0$.
 
The resulting time evolution operator $U_\mathrm{eff}$ for a single $\mathrm{Ni}^{2+}$ $S=1$ moment can then be written in the form (up to constants quadratic in $\mathcal{E}$)
\begin{widetext}\begin{align}
 	U_\mathrm{eff}(t_p,0) = &\mathds{1} + C^z_{A_1}(t_p) \mathcal{E}\cdot \mathcal{E}^\ast \left(S^z\right)^2 +C^z_{A_2}(t_p) \left(\iu \mathcal{E} \times \mathcal{E}^\ast\right) S^z \nonumber\\
 	&+ C^{xy}_E(t_p) \left[ \left(\mathcal{E}_x \mathcal{E}_y^\ast + \mathcal{E}_y \mathcal{E}_x^\ast  \right) \left(S^x S^y + S^y S^x\right) + \left(\mathcal{E}_x \mathcal{E}_x^\ast - \mathcal{E}_y \mathcal{E}_y^\ast \right) \left(\left(S^x\right)^2 - \left(S^y\right)^2 \right)  \right].\label{eq:ueff-final}
\end{align}
\end{widetext}
Here, we have labelled the coefficients by the irreducible representations of $D_3$ under which the respective electric field bilinears transform.
In general, these coefficients $C^{\alpha}_X= \Re C^\alpha_X + \iu \Im C^\alpha_X$ are complex, and become purely imaginary (real) in the off-resonant (resonant) limit.
We further stress that these coefficients $C^{\alpha}_X = C^{\alpha}_X(t_p)$ depend on the pump length $t_p$.

For technical details on the derivation of \eqref{eq:ueff-final} we refer the reader to App.~\ref{app:eval-eff-U}.

The first term in \eqref{eq:ueff-final} corresponds to a light-induced modulation of the single-ion anisotropy normal to the plane of incidence and is proportional to total intensity of the light, implying a polarization-independent result, with the coefficient 
\begin{equation}
	C^z_{A_1} = -\frac{\iu}{24}(2\mathsf{g}_0-3\mathsf{g}_1+\mathsf{g}_2)
\end{equation}
where $\mathsf{g}_j=\frac{1}{\omega-\epsilon_{m_j0}} \left[t_p - \frac{\eu^{\iu (\omega-\varepsilon_{m_j0})t_p}-1}{\iu (\omega- \varepsilon_{m_j0})} \right],\, j=0,1,2$, with $\epsilon_{m_j 0}$ the energy difference between the ground state and the excited state with $J_{\rm eff}=j$ in Eq.~\eqref{eq:H0m}. 

The matrix elements of the second term in \eqref{eq:ueff-final}, transforming in the $A_2$ irreducible representation, are proportional to the chiral intensity $\iu \mathcal{E} \times \mathcal{E}^\ast = \iu \mathcal{E}_\alpha \mathcal{E}_\beta^\ast \epsilon^{\alpha\beta}$ ($\alpha,\beta = x,y$) which is finite only for circularly polarized light, and couples to the $z$-component of the local moment.
In the off-resonant limit, this term is thus understood as a light-induced Zeeman magnetic field, which lies at the heart of the ``inverse Faraday effect'' .\cite{pershan66,kiri10,seiba19}
The corresponding coefficient reads
\begin{equation} \label{eq:CzA2}
	C^z_{A_2} = -\frac{\iu}{24}(2\mathsf{g}_0+3\mathsf{g}_1-5\mathsf{g}_2) 
\end{equation}

Further there are two terms in \eqref{eq:ueff-final} which belong to the two-dimensional $E$ representation of $D_3$ and are analogous to single-ion anisotropies within the $xy$-plane (which coincides with the crystallographic basal plane).
The coefficient reads
\begin{equation} \label{eq:CxyE}
	C^{xy}_{E} = \frac{\iu}{24}(2\mathsf{g}_0-3\mathsf{g}_1+\mathsf{g}_2)
\end{equation}

An inspection of the symmetry properties of the system reveals that in principle one can consider a second two-dimensional $E$ representation involving the spin bilinears $\{S^x,S^z\}$ and $\{S^y,S^z\}$, which however is absent from the $\mathcal{U}_\mathrm{eff}$.
This follows from a generalized dipole selection rule for the fictitious orbital angular momentum:
Noting that the dipole operator at normal incidence only involves $x$ and $y$ components, allowed transitions need to satisfy $\Delta J^z_\mathrm{eff} = \pm 1$, where it is understood that $J^z_\mathrm{eff} \equiv S^z$ in the $A_2$ ground state (in which the orbital angular momentum is quenched) and $J^z_\mathrm{eff} = l^z + S^z$ in the excited states.
It thus follows that matrix elements of $\mathcal{U}_\mathrm{eff}$ in the ground-state manifold derived from second-order perturbation theory are either diagonal or need to satisfy $\Delta J^z_\mathrm{eff} \equiv \Delta S^z = \pm 2$, which is only facilitated by products of two operators of the type $S^\pm = S^x \pm \iu S^y$, and thus $S^z$ cannot couple to any off-diagonal spin operators.

\begin{figure}[tb]
	\includegraphics[width=\columnwidth]{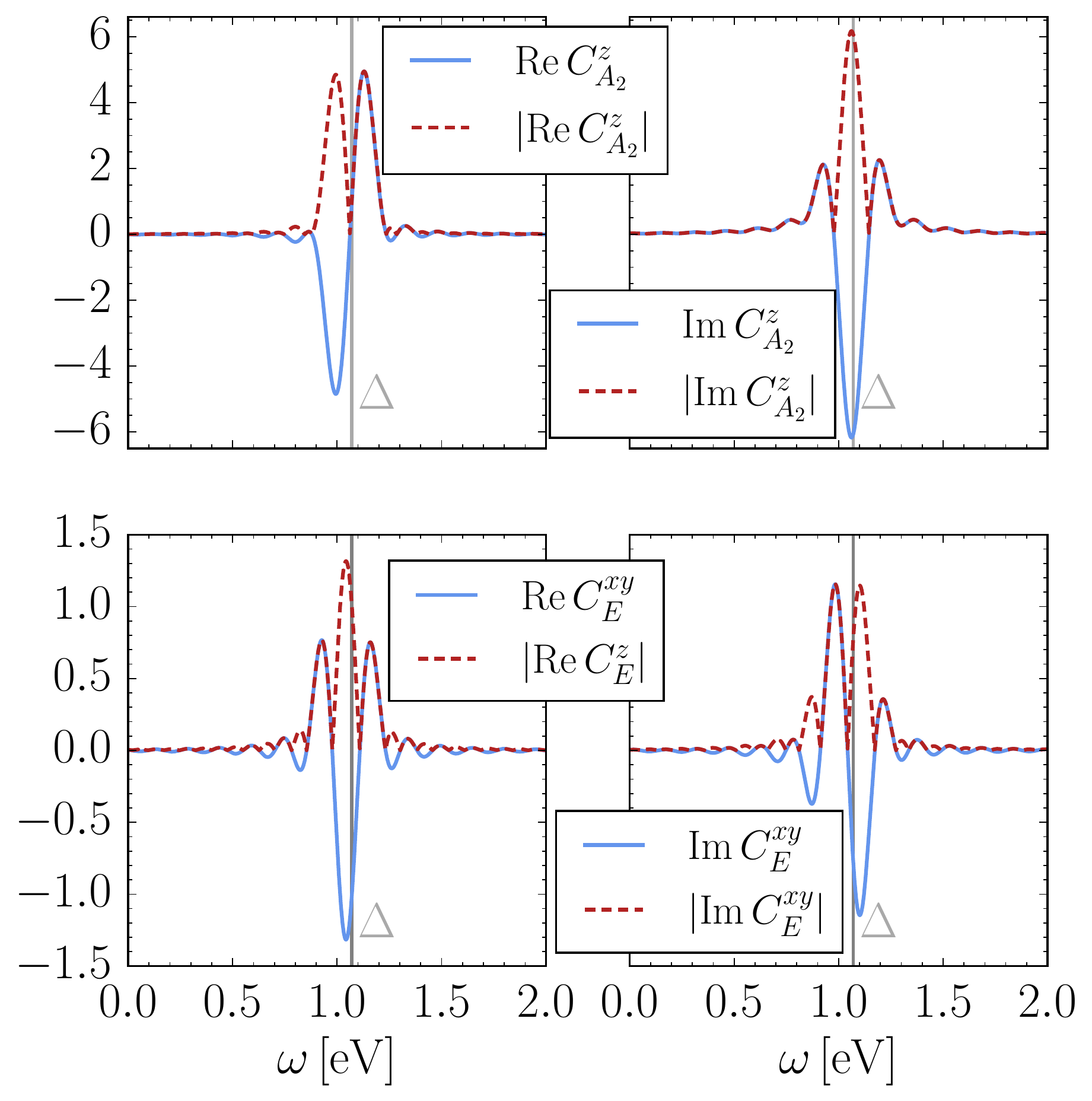}
	\caption{Plot of the real and imaginary parts of $C^z_{A_2}$ [Eq.~\eqref{eq:CzA2}] and $C^{xy}_E$ [Eq.~\eqref{eq:CxyE}] (and their absolute values for comparison with Ref.~\onlinecite{afanasiev20}) as a function of the light's driving frequency. Here, we have taken the crystal field-splitting\cite{afanasiev20} $\Delta = 1.07 \, \mathrm{eV}$ and spin-orbit coupling\cite{kang20} $\lambda = 0.08 \, \mathrm{eV}$, and a boxcar pump profile of $100\, \mathrm{fs} \approx 41.36 \, \mathrm{meV}$. Note a non-uniform pump pulse will likely lead to a further smearing of the frequency-dependence of the observed intensities. \label{fig:illu_cxy_cz}}
\end{figure}

We plot the frequency dependency of $C^z_{A_2}$ and $C^{xy}_E$, which are of relevance to the excitation of spin waves in $\nips$ (see also Sec.~\ref{sec:bulknips}), informed by realistic microscopic parameters, in Fig.~\ref{fig:illu_cxy_cz}. In particular, as we detail in Sec.~\ref{sec:bulknips}, when pumping with linearly polarized light, only the $C^{xy}_E$ term contributes to the in-plane mode $f_1$. On the other hand, exciting the out-of-plane mode $f_2$ requires spin bilinears $\{S^x,S^z\}$ and $\{S^y,S^z\}$, which is absent here due to the generalized dipole selection rule stated above. This explains the experimental observation in Ref.~\cite{afanasiev20} that only the $f_1$ mode is excited in the narrow resonant frequency range.


\section{Anisotropic exchange interactions and floquet spin Hamiltonian}
\label{sec:floquet}

Motivated by the observation of low energy out-of-plane mode $f_2$ in a broad frequency range transparent in optical absorption, we consider an off-resonant mechanism, where the Floquet spin Hamiltonian introduces effective magnetic fields that couple to the slow spin fields $\ve{n}, \ve{m}$ (see Eq.~\eqref{eq:nmEoM}).
Similar to the orbital resonance induced spin transition as discussed in Sec.~\ref{sec:atomic}, the SOC is essential here for the pump field to induce anisotropic spin interactions, and thus generate an effective staggered field along $\hat{z}$ or $\hat{y}$ which couples to the Néel vector.

To understand the possible spin interaction channels, we first obtain the effective bilinear spin Hamiltonian in equilibrium from a minimal two orbital Hubbard model on a honeycomb lattice. In particular, we will focus on the anisotropic spin exchanges which require SOC. In equilibrium, these anisotropic terms will modify the spin wave spectrum, but do not contribute to the effective magnetic fields when the zigzag order is the true ground state. Next, we obtain the corresponding non-equilibrium spin exchange through the Floquet formalism, and identify the effective field up to quadratic order in the pump electric field. 

\subsection{Equilibrium spin exchange interactions}

For simplicity, we ignore the ground state multiplet in $d^9 \underline{L}$ configuration, and restrict to the 4 states in the $d^8$ configuration at half-filling in the $e_g$ orbitals for each Ni$^{2+}$.
We will focus only on the nearest-neighbor (NN) and third-nearest-neighbor (TNN) terms, both of which are invariant under the two-fold rotation along the bond ($C'_2$), product of bond inversion (i.e. with the inversion center given by a midpoint of a bond), and mirror with the mirror plane perpendicular to the bond ($\mathcal{I}\times \sigma_v$), and time reversal (see also Fig.~\ref{fig:nips3_struct} for an illustration of the $C'_2$ axis and mirror plane). We then use the threefold rotational symmetry about a Ni site and translational symmetries to generate hopping terms for the full lattice.

We hence obtain the general symmetry allowed tight-binding Hamiltonian within $e_g$ orbitals (ignoring the ligand atoms), which reads
\begin{widetext}
\begin{align} \label{eq:ht}
\mathcal{H}_t=\sum_{\ver\in A,\ve{\delta}_i} \Psi^{\dagger}_{\ver} \left[W_{C_3}^{i-1}\left(w_0 \tau_0 \sigma_0+w_1\tau_z \sigma_0 +w_2 \tau_y \sigma_z + w_3 \tau_y \sigma_x\right) \left(W_{C_3}^{\dagger}\right)^{i-1} \right]\Psi_{\ver+\ve{\delta}_i}+\hc\ 
\end{align}
\end{widetext}
where $\Psi_{\ver}=\left( e_{\omega,\uparrow}, e_{\omega,\downarrow}, e_{\omega^2,\uparrow}, e_{\omega^2,\downarrow}\right)^\top_\ver$, $\ve{\delta}_i=\ve{\delta}^{(l)}_i$ with $ i=1,2,3$ denotes the NN ($l=1$) or TNN ($l=3$) bonds. $\ve{\delta}_1$ is along the $\hat{y}$ direction in trigonal coordinate, and $\ve{\delta}_i=\mathcal{R}_{C_3} {}^{i-1} \ve{\delta}_1$. Here, $W_{C_3}, \mathcal{R}_{C_3}$ are the three-fold rotation operations acting on the single-particle fermion operators $\Psi$ and 3d spatial vectors, respectively. 
$\tau_i, \sigma_i$ are identity ($i=0$) and Pauli matrices ($i=1,2,3$) in the orbital and spin space, respectively. Note that among the hopping integrals denoted as $w_\mu\, (\mu=0,...,3)$, $w_2$ and $ w_3$ require SOC.

To capture the onsite interactions, for simplicity, we consider only the intra-orbital Hubbard repulsion $\mathcal{H}_U=U \sum_{i,\alpha} \hat{n}_{i,\alpha} (\hat{n}_{i,\alpha}-1)$, where $\alpha$ labels the $e_g$ orbitals. Note that Hund's coupling ($J_H$) and inter-orbital Hubbard interaction ($U'$) do not contribute to the exchange coupling at the leading order in $J_H, U'$ because the atoms active in the excited state perturbed by $\mathcal{H}_t$ are occupied by one electron (hole). The effective spin Hamiltonian at half-filling obtained through second-order perturbation theory in $t/U \ll 1$ is given by
\begin{align}
\mathcal{H}_{\rm eff}^{\rm ex}= - \mathcal{P}_s \mathcal{H}_t \mathcal{P}_d \frac{1}{\mathcal{H}_U} \mathcal{P}_d \mathcal{H}_t \mathcal{P}_s,
\end{align} 
where $\mathcal{P}_{s,d}$ are projection operators on the single and double electron occupancy space. $\mathcal{H}_{\rm eff}^{\rm ex}$ up to a constant is
\begin{align}\label{eq:effspin}
\mathcal{H}_{\rm eff}^{\rm ex}=\sum_{\ve{r},\ve{\delta}_i} \vec{S}_{\ve{r}}^\top \mathsf{\Gamma}^{\ve{\delta}_i}
\vec{S}_{\ve{r}+\ve{\delta}_i}
\end{align}
where $\mathsf{\Gamma}^{\ve{\delta}_i}=\mathcal{R}_{C_3}^{i-1} \mathsf{\Gamma}^{\ve{\delta}_1} \mathcal{R}_{C_3}^\top {}^{i-1}$, and
\begin{align}\label{eq:effspin1}
\mathsf{\Gamma}^{\ve{\delta}_1}&=
\begin{pmatrix}
J_l-J_l^{\prime}+J_l^{\prime\prime} & 0 & J_{l,xz} \\
0 & J_l-J_l^{\prime}-J_l^{\prime\prime} & 0\\
J_{l,xz} & 0 & J_l+J_l^{\prime}-J_l^{\prime\prime} \\
\end{pmatrix}.
\end{align}
 Here, $J_l$ is the $l$-th NN isotropic Heisenberg exchange, $J_l^\prime, J_l^{\prime\prime} , J_{l,xz}$ are the anisotropic spin exchange that requires SOC. Specifically, we find 
\begin{align}
J_{l} & = \frac{2(w_0^2 + w_1 ^2)}{U},\non\\
J_{l}^\prime & = \frac{-2 w_2^2}{U},\, J_{i}^{\prime\prime}=\frac{2 w_3^2}{U},\,
J_{l,xz} = \frac{4 w_2 w_3}{U},
\end{align}
with $w_\mu = w_\mu^{(l)}$ denoting the $\mu = 0, \dots, 3$ hopping amplitudes on $l=1,3$-nearest neighbor bonds as introduced in \eqref{eq:ht}.

In passing, we note that there are a few cautions to fully apply the minimal model  to $\nips$:

\emph{First}, in this consideration, the exchange paths are limited to only between Ni ions. However, in transition metal trichalcogenide, the ligand atoms play an important role to determine the sign of the exchange interactions.
Indeed, the NN Heisenberg term is found to be ferromagnetic~\cite{gu19}, which cannot be explained by the anti-ferromagnetic superexchange presented in Eq.~\eqref{eq:effspin1}.
It instead requires considering the super-exchange paths through the ligand atoms following the Goodenough Kanamori approach~\cite{Flem82,Wildes15}.
Similarly, for the TNN spin exchange, the super-superexchange paths are dominant~\cite{gu19,Chaudhary2020}.
Nevertheless, the current consideration obtains all symmetry allowed spin exchange terms and will be taken to derive the Floquet spin Hamiltonian with the right symmetry.

\emph{Second}, besides electron hopping $w_2, w_3$, SOC may also enter into the effective spin Hamiltonian through on-site term $\mathcal{H}_{\rm SOC, onsite}= \lambda_{\mathrm{SOC}}\sum_i \Psi^{\dagger}_i \tau_y \sigma_z \Psi_i$.
However, we note that the ground state multiplet is a orbital singlet, $\mathcal{H}_{\rm SOC, onsite}$ does not have non-zero matrix element within the ground state multiplet manifold.
A careful analysis including both $\mathcal{H}_t$ and $\mathcal{H}_{\rm onsite, SOC}$ indicates that the onsite SOC does not contribute to the anisotropic spin Hamiltonian up to $\frac{w_0^2 \lambda^2}{U^3}$, so will be ignored in further discussions. 

\subsection{Floquet spin Hamiltonian}
We consider the following periodically driven Hubbard model
\begin{align}
\mathcal{H}_t(t) &= \sum_{\langle i, j \rangle} w_{i s_1, j s_2}^{\alpha\beta} \eu^{-\iu e \ve{A}(t)\cdot \ve{r}_{ij}} \hat{v}_{i \alpha s_1, j \beta s_2}+\hc \non\\
\mathcal{H}_U &= U\sum_{i} \hat{n}_{i\alpha} (\hat{n}_{i\alpha}-1)
\end{align}
where $\ve{r}_{ij}=\ve{r}_i - \ve{r}_j$, and the effects of the pump laser is manifested in the kinetic energy $\mathcal{H}_t(t)$ through the Peierls substitution. For compactness, we have introduced the electron hopping operator $\hat{v}_{i \alpha s, j \beta s'}=e^{\dagger}_{i\alpha s} e_{j \beta s'}$ that denotes the hopping of spin $s'$ orbital $\beta$ electron at site $j$ to spin $s$ orbital $\alpha$ electron at site $i$.

Consider the electric field $\vec{E}=\frac{1}{2} \left(\vec{\mathcal{E}} \eu^{\iu \omega t}+\vec{\mathcal{E}}^* \eu^{-\iu \omega t}\right)$, the gauge term reads
\begin{align}
\eu^{-\iu e \ve{A}(t)\cdot \ve{r}_{ij}}= \eu^{\iu \frac{\Omega_{ij}}{\omega} \sin \left(\omega t +\phi_{ij}\right)} 
= \sum_{n=-\infty}^{n=\infty} \eu^{\iu n (\omega t +\phi_{ij})}\mathcal{J}_n (\frac{\Omega_{ij}}{\omega})
\end{align}  
where the ``Rabi'' frequency of the laser for bond $\langle i j \rangle$ is $\Omega_{ij}=e \sqrt{(\re \Ef \cdot \ve{r}_{ij} )^2+(\im \Ef \cdot \ve{r}_{ij} )^2}$, the phase $\phi_{ij}$ is determined by $\tan \phi_{ij}=\im \Ef \cdot \ve{r}_{ij}/\re \Ef \cdot \ve{r}_{ij}$
In the off-resonant limit, i.e.\ when $|U-n \omega|$ is much greater than the band-width of the many-body excited states, we can ignoring the heating due to, e.g.\ doublon decay~\cite{Hejazi2019}, the Floquet spin Hamiltonian from the Hubbard model can be expressed as
\begin{widetext}
\begin{align}
\mathcal{H}_{\rm eff}^{\rm ex} =&- \sum_{\langle i j\rangle, n,n' }w^{\alpha' \beta'}_{j s'_1, i s'_2 }  w^{\alpha \beta}_{i s_1, j s_2 } \mathcal{J}_{n'}(\frac{\Omega_{ji}}{\omega}) \mathcal{J}_n (\frac{\Omega_{ij}}{\omega}) \eu^{\iu (n+n')\omega t} \eu^{\iu (n' \phi_{ji}+n \phi_{ij})}\mathcal{P}_s \hat{v}_{j \alpha' s'_1, i \beta' s'_2} \frac{\mathcal{P}_d}{\mathcal{H}_U+ n \omega}\hat{v}_{i \alpha s_1, j \beta s_2} \mathcal{P}_s \non\\
\rightarrow & - \sum_{\langle i j\rangle,n }w^{\alpha' \beta'}_{j s'_1, i s'_2 }  w^{\alpha \beta}_{i s_1, j s_2 } \mathcal{J}_{n}(\frac{\Omega_{ji}}{\omega}) \mathcal{J}_{-n} (\frac{\Omega_{ij}}{\omega}) \eu^{\iu n ( \phi_{ji}- \phi_{ij})}\mathcal{P}_s \hat{v}_{j \alpha' s'_1, i \beta' s'_2} \frac{\mathcal{P}_d}{\mathcal{H}_U-n \omega}\hat{v}_{i \alpha s_1, j \beta s_2} \mathcal{P}_s \non\\
=& - \sum_{\langle i j\rangle,n }w^{\alpha' \beta'}_{j s'_1, i s'_2 }  w^{\alpha \beta}_{i s_1, j s_2 } \mathcal{J}^2_{n}(\frac{\Omega_{ji}}{\omega})\, \mathcal{P}_s \hat{v}_{j \alpha' s'_1, i \beta' s'_2} \frac{\mathcal{P}_d}{\mathcal{H}_U-n \omega}\hat{v}_{i \alpha s_1, j \beta s_2} \mathcal{P}_s 
\label{eq:heff_ex}
\end{align}
\end{widetext}
From the first to second line, we have considered only the time-independent effective spin Hamiltonian, and to obtain the last line, we have used $\phi_{ij}=\phi_{ji}+\pi$. Note that compared with the equilibrium spin Hamiltonian, the pump laser only modifies the effective exchange coupling strength of a given bond $\langle i j \rangle$ by a factor $\mathcal{J}^2_{n}(\frac{\Omega_{ji}}{\omega}) \frac{U}{U-n\omega}$. For linearly polarized light, $\Omega_{ij}=e |\re \Ef\cdot \ve{r}_{ij}|$. Importantly, $\Omega_{\ver, \ver+\ve{\delta}_i^{(l)}}$ for the three $l=1,3$-nearest neighbor bonds related by $C_3$ rotation can be different, which may lead to anisotropy terms taking the N\'eel vector out of the equilibrium ground state manifold. On the other hand, for circularly polarized light, as $\re \Ef \perp \im \Ef$ and $|\re \Ef |=|\im \Ef|$, we find $\Omega_{\ver, \ver+\ve{\delta}_i}$ the same for $i=1,2,3$. As a result, the circularly polarized pump laser could only modify the strength of the spin exchange while preserving the $C_3$ rotational symmetry, which may significantly modify the ground state manifold when the equilibrium spin system is near a phase transition.

\section{Application: Magnon dynamics in bulk $\nips{}$} \label{sec:bulknips}

\subsection{Equilibrium spin Hamiltonian} 
Both experimental and first-principle studies have shown that the effective spin Hamiltonian for $\nips$ can be captured by the Heisenberg model with NN, SNN and TNN spin exchange. As the ground state multiplet for a Ni$^{2+}$ ion is spin triplet and orbital singlet, the effective spin orbit coupling (SOC) on the ground state manifold is small, and requires mixing with higher excited levels. In the following, we express the equilibrium spin Hamiltonian as $\mathcal{H}_{\rm spin}=\mathcal{H}_{\rm spin}^{(0)}+\mathcal{H}_{\rm spin}^{(1)}$, where only $\mathcal{H}_{\rm spin}^{(1)}$ requires SOC. 

The zigzag N\'eel order is stabilized by the isotropic Heisenberg terms~\cite{HuJ2019}
\begin{align}\label{eq:Hspin0}
\mathcal{H}_{\rm spin}^{(0)}=J_1\sum_{\langle ij\rangle_{1}} \vec{S}_i\cdot \vec{S}_j+J_2\sum_{\langle ij\rangle_{2}} \vec{S}_i\cdot \vec{S}_j+J_3\sum_{\langle ij\rangle_{3}} \vec{S}_i\cdot \vec{S}_j
\end{align}
with $\langle i j \rangle_l$ denotes the $l$-nearest neighbor bond, $|J_3|\gg |J_1|\gg |J_2|$ and $J_3>0, J_1<0$. The strength of $J_3$ is found at order 10meV. 

$\mathcal{H}_{\rm spin}^{(1)}$ are generally weaker in magnitude, but they are important in the observation of coherent magnon oscillation at terahertz frequency. In the following, we model the spin anisotropy by the single ion anisotropy~\cite{Joy1992}, which reads
\begin{align}\label{eq:Hspin1}
\mathcal{H}_{\rm spin}^{(1)}=D_{z} \sum_i S_{i,z}^2+D_{xy} \sum_i \left(S_{i,y}^2-S_{i,x}^2\right),
\end{align}
and $|D_z| \gg |D_{xy}|$. Note that with ${\rm D}_{3d}$ point group symmetry of the crystal, only $D_z\neq 0$ is allowed. However, the bulk $\nips$ crystalizes in a monoclinic structure~\cite{afanasiev20}, which breaks the three-fold rotation due to displacement of adjacent layers along the $\hat{a}$-axis. This allows for a much weaker $D_{xy}\neq 0$. 


In passing, we note that other anisotropic spin exchange terms may also enter into $\mathcal{H}^{(1)}_{\rm spin}$. However, as we discuss below, they do not modify the dynamics of slow modes qualitatively. 

\subsection{Slow modes and low-energy equation of motion}
Here, we consider an easy-plane zigzag order ($D_z>0$) which favors alignment along $\vhat{x}$ ($D_{xy}>0$). 
The slow modes, i.e.\ the Goldstone modes, are the transverse fluctuations of the N\'eel order, whose energy gap are determined by the the anisotropy terms in Eq.~\eqref{eq:Hspin1}.
When the spatial variations of the ordered state is small and slowly varying, the semiclassical spin configuration at site $\ver$ can be described as
\begin{align}
\vec{S}_\ver \equiv \vec{S}_{{\bf R},\alpha}=(-1)^\alpha \eu^{i {\bf M}\cdot {\bf R}} \,S\bm{n}_{\ver} \sqrt{1-(\bm{m}_{\ver})^2}+S\bm{m}_{\ver},
\label{eq:NLSMN}
\end{align}
where ${\bf R},\alpha$ denote the unit cell coordinate and sublattice label of the site $\ver$, $\bm{n}_\ver=(n_0, n_y, n_z)_\ver$ denotes the staggered component of the spin fields in terms of the static order parameter $n_0 \vhat{x}$ and transverse fluctuations $n_y \vhat{y}, n_z \vhat{z}$, and $\bm{m}_\ver$ is the ferromagnetic component of the fluctuating spin fields. The phase factor $(-1)^\alpha \eu^{i {\bf M}\cdot {\bf R}} $ is chosen to describe the zigzag ordering pattern (see detailed discussions in Appendix~\ref{app:EoM}). 

We also include a general form of the {\it effective Hamiltonian} to model the effect of the pump field in the ultrafast regime,
\begin{align} \label{eq:heff_pump_bulk}
\mathcal{H}_{\rm eff}^{\rm pump}= - S( \ve{h}_n \cdot \ve{n} +  \ve{h}_m \cdot \ve{m} ),
\end{align}
where $\ve{h}_{n,m}$ are the effective fields to be determined microscopically up to quadratic order in $|\ve{E}|^2$. 

From Eqs.~\eqref{eq:Hspin0},~\eqref{eq:Hspin1},~\eqref{eq:NLSMN} and~\eqref{eq:heff_pump_bulk}, we arrive at the equation of motion for the spatially homogeneous slow modes:
\begin{align}\label{eq:EoMnm}
\dot{n_y}&=\chi^{-1} m_z - h_{m,z}, \quad \dot{m_z}= - \kappa_{n_y}  n_y + h_{n,y};\non\\
\dot{n_z}&=-\chi^{-1} m_y + h_{m,y}, \quad \dot{m_y}=  \kappa_{n_z}  n_z - h_{n,z}.
\end{align}
Here, $\chi^{-1} \sim J S $ is the uniform spin susceptibility, $\kappa_{n_y}, \kappa_{n_z} \sim D S$ are the anisotropy. Note that $n_y, m_z$ and $n_z, m_y$ are two sets of conjugate fields. In the probe field period, they form two sets of harmonic oscillators with frequencies $\Omega_{f_1}=\Omega_{n_y}=\sqrt{\kappa_{n_y} \chi^{-1}}$, $\Omega_{f_2}=\Omega_{n_z}=\sqrt{\kappa_{n_z}\chi^{-1}}$. In the pump period starting at $t=0$ of duration $t_p$, assuming square pulses $\ve{h}_{n,m}=\bar{\ve{h}}_{n,m} (\Theta(t)-\Theta(t-t_p))$, the effective magnetic fields  exert forces to the spin fields, which determine the initial condition of the free oscillation in the probe period:
\begin{align}
m_z (t_p^+) &= \bar{h}_{n,y} t_p ,\quad \partial_t m_z(t_p^+)= \kappa_{n_y}\bar{h}_{m,z} t_p \,; \non\\
m_y (t_p^+) &= -\bar{h}_{n,z} t_p ,\quad \partial_t m_z(t_p^+)= \kappa_{n_z}\bar{h}_{m,y} t_p. \label{eq:initial-cond-fields}
\end{align}
\subsection{Pump-induced effective fields and initial conditions}

\subsubsection{Pumping the orbital resonance}

First, we discuss the initial conditions for the magnetization and the N\'eel vector due to the effective light-induced single-ion anisotropy.
While the limit of off-resonant driving (with $t_p (\omega- \varepsilon_{m0}) \gg 1$) is readily treated in the framework presented in the previous subsection by rewriting the pump-induced effective Hamiltonian in terms of the magnetization and Néel vector continuum fields and then identifying the linear terms (with their coefficients constituting source fields), this treatment fails for driving frequencies near an orbital resonance for which the effective time evolution in the $S=1$ ground-state sector is no longer unitary, as shown in \ref{sec:tdpt}. Further, we stress that the quadratic time dependence in \eqref{eq:UI_analytical} implies that a light pulse with a time-independent fluence leads to an explicit time-dependence of any effective pump-induced low-energy source term, rendering the assumption leading to \eqref{eq:initial-cond-fields} invalid.

To mitigate this, we instead directly compute the initial conditions for the magnetization and Néel order parameter fields through evaluating the respective time-evolved microscopic defining spin expectation values.
The magnetization is given by $m_z(t_p) = \frac{1}{N} \sum_i \braket{n_0| S^z_i | n_0}(t_p)$, where the expectation value is taken with respect to the Néel reference state $\ket{n_0} = \prod_{i} \ket{S^x=(-1)^{s_i} \eu^{\iu \mathbf{M} \cdot \mathbf{R}_i}}_i$ corresponding to $S=1$ moments fully polarized in $\pm \hat{x}$ direction depending on unit cell coordinate $\mathbf{R}_i$ and sublattice $\alpha_i$ of site $i$ (see also the parametrization \eqref{eq:NLSMN}).
Analogously, the $\hat{y}$-component of the Néel order parameter is obtained as $n_y(t_p) =\frac{1}{N} \sum_{i}  (-1)^{\alpha_i} \eu^{\iu \mathbf{M} \cdot \mathbf{R}_i} \braket{n_0 | S^y_i | n_0}(t_p)$.
Using the time-evolution operator in the interaction picture, we thus find for the expectation value at site $i$
\begin{equation}
    \braket{n_0 | S^z_i | n_0}(t_p) = \braket{\pm_i | U_\mathrm{eff}^\dagger(t_p,0) S^z U_\mathrm{eff}(t_p,0) | \pm_i},
\end{equation}
With the general form of \eqref{eq:ueff-final} we can write to second order
\begin{align} \label{eq:Ut_s_u}
    &U_\mathrm{eff}^\dagger(t_p,0) S^z U_\mathrm{eff}(t_p,0) \approx S^z \nonumber\\&+\sum_\mathcal{O} \left( \Re (C_\mathcal{O}(\mathcal{E},\mathcal{E}^\ast) \left\{S^z, \mathcal{O} \right\} + \iu \Im( C_\mathcal{O}(\mathcal{E},\mathcal{E}^\ast)) \left[S^z,\mathcal{O} \right] \right),
\end{align}
where $\mathcal{O}$ denote the (products of) spin operators appearing in \eqref{eq:ueff-final}, and $C_\mathcal{O}(\mathcal{E})$ their respective coefficients (here, we have absorbed the electric field bilinears into the coefficients for ease of notation).
We note that the non-unitarity of time evolution is reflected in the presence of the anticommutator, and unitarity is recovered in the off-resonant limit where $\Re C_\mathcal{O} = 0$, as argued earlier.
Taking the expectation value of \eqref{eq:Ut_s_u} with respect to $\ket{S^x = \pm}$, we find that the only non-vanishing contributions are given by $\braket{\{S^z,S^z\}} = 1$ and $\braket{[S^z,S^x S^y+ S^y S^x]} = -\iu$, as well as $\braket{\pm|\{S^y,S^x S^y+ S^y S^x\}|\pm} = \pm 1$ and $\braket{\pm|[S^y,S^z]|\pm} = \pm \iu$.
We hence find the initial conditions for the equilibrium relaxational dynamics of the out-of-plane magnetization and its velocity induced by the perturbation as
\begin{subequations}
\begin{align} \label{eq:initial-cond-reso-a}
    m_z(t_p^+) &= \Re(C^{z}_{A_2}) \iu \mathcal{E} \times \mathcal{E}^\ast + \Im(C^{xy}_E) (\mathcal{E}_x \mathcal{E}_y^\ast + \hc) \\
    \partial_t m_z(t_p^+) &= - \kappa n_y(t_p^+) \nonumber\\
    &= -\kappa \Re(C^{xy}_{E}) (\mathcal{E}_x \mathcal{E}_y^\ast + \hc) + \kappa \Im(C^{z}_{A_2}) \iu \mathcal{E} \times \mathcal{E}^\ast, \label{eq:initial-cond-reso-b}
\end{align}
\end{subequations}
which supersede \eqref{eq:initial-cond-fields} in the regime where the coupling of the light to the orbital resonance is most dominant.
We stress that Eqs.~\eqref{eq:initial-cond-reso-a} and \eqref{eq:initial-cond-reso-b} have multiple qualitative/semi-quantitative implications for experiment. Away for finite pump lengths (away from the resonant or off-resonant limits), both $C^z_{A_2}$ and $C^{xy}_E$ are complex.
Hence, by pumping with linearly or circularly polarized light the first (second) term in \eqref{eq:initial-cond-reso-a} and \eqref{eq:initial-cond-reso-b} are activated and determine the initial conditions for the relaxational dynamics.
By tuning $\omega$ (e.g. to the off-resonant limit such that the coefficients become purely imaginary), the balance of \eqref{eq:initial-cond-reso-a} and \eqref{eq:initial-cond-reso-b} is changed (and thus oscillations become more $\sin$- or $\cos$-like) in a characteristic manner, which can be observed by tracking the phase of the magnetization oscillations with respect to a fixed time $t_0=0$ (assuming constant pump lengths).

\subsubsection{Floquet modification of spin exchange}

Next, we obtain the effective magnetic field through the bond-dependent anisotropic Floquet spin Hamiltonian, which is induced by the linearly polarized pump. Note the staggered field along $\hat{z}$ and $\hat{y}$, i.e.\ $h_{n_z}$ and $h_{n_y}$, can be obtained from $S_\ver^x S_{\ver+\ve{\delta}_i}^z +S_\ver^z S_{\ver+\ve{\delta}_i}^x$ and $S_\ver^x S_{\ver+\ve{\delta}_i}^y +S_\ver^y S_{\ver+\ve{\delta}_i}^x$ types of exchanges, respectively. To be specific, defining $J_{x z,\ve{\delta}_i}^{(n)}$ as the exchange coupling coefficient for spin bilinear $\{S^x, S^z\}$ on bond $\langle \ver,\ver+\ve{\delta}_i\rangle$ from the n-th order photon absorption, i.e.\ with resolvent $\frac{1}{U-n\omega}$, we find 
\begin{align}
\mathcal{H}_{{\rm eff},n_z}^{\rm pump}=&\sum_{\ver, i} J_{x z,\ve{\delta}_i}^{(1)} \left( \langle S_{\ver}^x \rangle S_{\ver+\ve{\delta}_i}^z + S_{\ver}^z \langle S_{\ver+\ve{\delta}_i}^x\rangle \right) + \mathcal{O}(|\Ef|^4)\non\\
=& -2 S^2 n_0 \sum_{\ver, i} J_{x z,\ve{\delta}_i}^{(1)} n_z\non\\
 =&-\frac{2 S^2 n_0}{\uc} \int \du^2 \ve{x} \left[\sum_i J_{x z,\ve{\delta}_i}^{(1)}\right] n_z.
 \label{eq:heff_pump_nz}
\end{align}
From Eqs.~\eqref{eq:heff_ex} and~\eqref{eq:heff_pump_nz}, we can read off the effective staggered magnetic field which couples to the $\hat{z}$-component of the N\'eel vector as
\begin{align}
h_{n,z} &=- 2 n_0 S \left[\sum_i J_{x z,\ve{\delta}_i}^{(1)}\right] \non\\
&=- \frac{8 n_0 w_2 w_3 S}{U-\omega}\left[ \left(\frac{\Omega_{\ve{\delta}_1}}{\omega}\right)^2-\frac{1}{2} \left(\frac{\Omega_{\ve{\delta}_2}}{\omega}\right)^2 -\frac{1}{2} \left(\frac{\Omega_{\ve{\delta}_3}}{\omega}\right)^2\right]\non\\
&= \frac{6 n_0 w_2 w_3 S}{U-\omega}  \left(\frac{\Omega_{\rm rabi}}{\omega}\right)^2 \cos 2\phi 
\end{align}
where $\Omega_{\rm rabi}= e |\re \Ef| |\ve{\delta}_i|$.
Similarly we obtain the field for the in-plane component, 
\begin{align}
h_{n,y} &=- 2 n_0 S \left[\sum_i J_{x y,\ve{\delta}_i}^{(1)}\right]= -\frac{3 n_0 w_3^2 S}{U-\omega}  \left(\frac{\Omega_{\rm rabi}}{\omega}\right)^2 \sin 2\phi .
\end{align}

Using Eqs.~\eqref{eq:initial-cond-fields} that are valid in the off-resonant limit, we can find the initial conditions for the magnetization oscillations:
\begin{align}
m_z (t_p^+) &\sim -\frac{w_3^2 t_p }{U-\omega}  \left(\frac{\Omega_{\rm rabi}}{\omega}\right)^2 \sin 2\phi ,\quad \partial_t m_z(t_p^+)= 0 \,; \non\\
m_y (t_p^+) &\sim -\frac{w_2 w_3 t_p}{U-\omega}  \left(\frac{\Omega_{\rm rabi}}{\omega}\right)^2 \cos 2\phi ,\quad \partial_t m_z(t_p^+)= 0. \label{eq:initial-cond-ex}
\end{align}


\subsection{Experimental consequences}
We discuss implications from our theory to the experiment carried out by Afanasiev et al.\ in Ref.~\onlinecite{afanasiev20}, where the pump beam is linearly polarized. 

From Eqs.~\eqref{eq:initial-cond-reso-a} and ~\eqref{eq:initial-cond-reso-b}, the resonant mechanism we proposed through pumping orbital resonances excites {\it only} the in-plane $f_1$ mode (i.e.\ out-of-plane uniform magnetization).
The out-of-plane $f_2$ mode (i.e.\ in-plane uniform magnetization) cannot be excited due to the absence of the single-ion spin bilinear $\{S^x, S^z\}$ as a consequence of the orbital selection rule (see Sec.~\ref{sec:tdpt} for detailed discussions).
The amplitude of the magnetization oscillation $\bar{m}_z$ is determined by the initial conditions of $m_z(t_p^{+})$ from Eqs.~\eqref{eq:initial-cond-reso-a} and ~\eqref{eq:initial-cond-reso-b}, and reads
\begin{align}
 \bar{m}_z &= \left(m_z(t_p^+)^2+\left(\frac{\partial_t m_z(t_p^+)}{\omega}\right)^2 \right)^{1/2}\non\\
 & = \left(\im (C^{xy}_E)^2+ \frac{\kappa_{n_y}}{\chi^{-1}} \re (C^{xy}_E)^2 \right)^{1/2}.
\end{align}
Note that, since $\im C^{xy}_E$ and $\re C^{xy}_E$ have zeros at different $\omega$ (see also Fig.~\ref{fig:illu_cxy_cz}), above form implies that the experimentally observed amplitude does not feature zeros as a function of $\omega$, however in our case $\kappa/\chi^{-1} \sim D/J$ is small so that we expect the contribution from $\im (C^{xy}_E)$ to strongly dominate and determine the frequency dependence of $\bar{m}_z$.

From Eq.~\eqref{eq:initial-cond-ex}, the off-resonant mechanism we proposed through anisotropic Floquet spin exchange can excite both $f_1$ and $f_2$ modes. Thus, according to our theory, the observation of {\it only} $f_2$ mode in the optical transparent photon energy window $0.1\,{\text{eV}}\sim 0.9\,{\text{eV}}$ implies that the hopping integrals satisfy $w_2 \gg w_3$.

We note that our theory shows that the $f_1$ mode can also be pumped using circularly polarized light via the orbital-resonance mechanism, according to Eqs.~\eqref{eq:initial-cond-reso-a} and ~\eqref{eq:initial-cond-reso-b}, constituting an example of the previously discussed inverse Faraday effect \cite{pershan66}.

\section{Application: Dynamics of spin waves and vortices in monolayer $\nips{}$}
\label{sec:app_bkt}

In this section, we propose the use of pump-probe methods to study the collective dynamics spanning three distinct phases of two-dimensional XY-like magnets, arguing that the technique has the capability to observe effects associated with spin waves, vortices, and quasi-long-range order.

This possibility is built on remarkable progress made in recent years exfoliating van der Waals materials which remain magnetically ordered down to the monolayer limit \cite{mak2019,gibertini20}.
We therefore now turn to a single (two-dimensional) layer of $\nips$ for which recent Raman-scattering experiments find no signature of magnetic ordering in the monolayer limit \cite{kim19}.
Given the dominant easy-plane anisotropy in (bulk) $\nips$ as discussed in Sec.~\ref{sec:bulknips}, these findings suggest the intriguing possibility that the interaction between the in-plane local moments has an (approximate) $\SO(2)$-symmetry and thus monolayer $\nips$ may realize an XY model.
In two-dimensional XY models fluctuations prevent the spontaneous breaking of the $\SO(2)$ symmetry, and the model instead shows a topological phase transition at $T_\mathrm{BKT}$ from a regime with algebraically decaying correlations (at low temperatures $T < T_\mathrm{BKT}$) to an exponential decay (high temperatures $T > T_\mathrm{BKT}$) driven by the proliferation of vortices as topological defects of the order-parameter field, as shown by Berezinskii, Kosterlitz and Thouless \cite{berezinsky70,kosterlitz73}.
We note that surface termination effects have an impact on the precise nature and strength of interactions and anisotropies, and thus may lead to the distinct behaviours of monolayer and few-layer systems, while a naive model for the latter in terms of weakly coupled XY models would also predict the absence of long-range order.
A detailed understanding requires a realistic ab-initio modeling of the microscopic interactions which we leave for further studies.

As established experimental techniques for resolving the dynamics of magnets, e.g.\ neutron scattering, are only of limited applicability to 2D van der Waals magnets, the purpose of the following section is to study how the unconventional dynamics of BKT physics can be resolved within the ultrafast optical pump-probe framework outlined in Sec.~\ref{sec:framework}.

\begin{figure}[tb]
    \centering
    \includegraphics[width=\columnwidth]{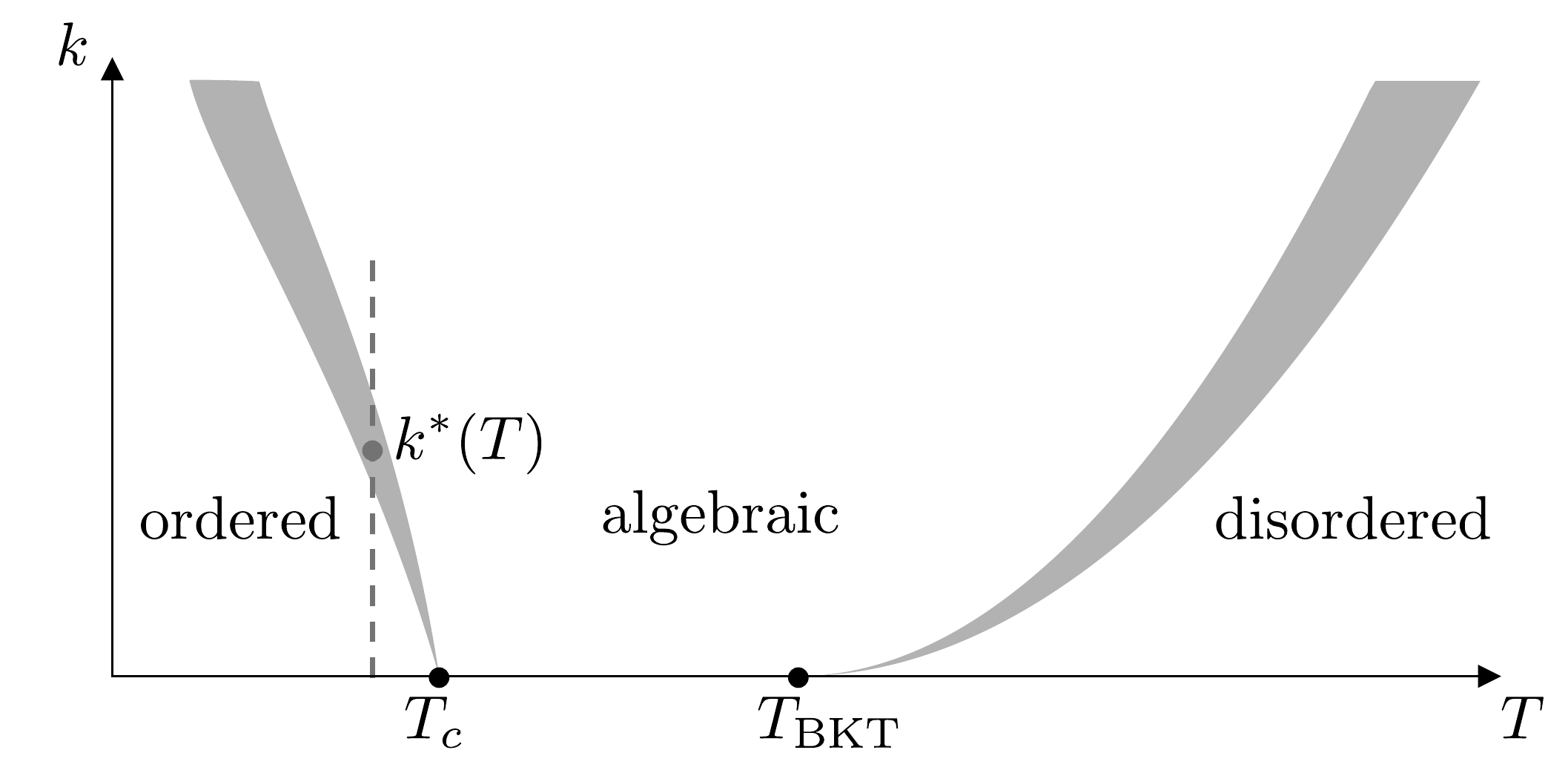}
    \caption{Schematic illustration of the phases of the XY model with a 6-fold anisotropy as a function of temperature $T$ and at momenta $k$. The grey areas denote crossovers at finite momentum $k$.
    A given temperature $T$ within the ordered phase $T < T_c$ is seen to define a critical momentum $k^\ast$ above which the dynamics is well-described in terms of the critical BKT scaling.}
    \label{fig:k-vs-T}
\end{figure}

We obtain a minimal action based on above considerations by parametrizing the in-plane Néel order parameter $\ve{n} = n_0 \left(\cos \phi, \sin \phi,0\right)$ and the magnetization $\ve{m} = m_z \hat{z}$, obtaining
\begin{align} \label{eq:s_xy}
	\mathcal{S}_\mathrm{XY} &= \frac{1}{2 v_\mathrm{u.c.}} \int \du t \du^2 \ve{x} \big\{ 2 S n_0^2 m_z \partial_t \phi \nonumber\\ &- \left[\rho n_0^2 (\nabla \phi)^2 + m_\mathrm{eff}^2 S^2 m_z^2 \right]  \big\},
\end{align}
where it becomes clear from the first term that the in-plane angle of the Néel order parameter, $\phi$, and the transverse magnetization $m_z$ are conjugate variables, for which \emph{linear} equations of motion can be derived. $\rho\sim J S^2$ is the spin stiffness, $m_\mathrm{eff}^2\sim J$ is given by the uniform spin susceptibility.
We further allow for a (weak) in-plane anisotropy of strength $h_p$ compatible with the point group symmetry of the system ($D_{3d}$), which can be written as
\begin{equation} \label{eq:s-p}
	\mathcal{S}_p = \frac{1}{2 v_\mathrm{u.c.}} \int \du t \du^2 \ve{x} \big\{ 2 h_p \cos\left(p \phi\right)\},
\end{equation}
where $p=6$ for the zigzag Néel order on the honeycomb lattice with spin-orbit-coupling as discussed previously. Renormalization group studies~\cite{Jose1977} found two phase transitions, at $T_{\rm BKT}$ and $T_{\rm c}$ when $p=6$, where $T_{\rm BKT}>T_{\rm c}$.
Below $T_{\rm c}$, the system orders into one of the six degenerate minima selected by the anisotropic term $\mathcal{S}_p$. Above $T_{\rm BKT}$, the system is disordered with exponential decaying correlations.
At any $T_{\rm c}<T<T_{\rm BKT}$, the system exhibits critical algebraic order.
The phase diagram as a function of temperature (as well as scaling of excited states at finite momentum) is illustrated in Fig.~\ref{fig:k-vs-T}.
Following the previously described framework, we first model the pump excitation through ``source'' terms which couple linearly to the low-energy degrees of freedom for the duration of the pump pulse, and thus take the system out of equilibrium.
The subsequent relaxational dynamics then proceeds according to the equilibrium equations of motion, with initial conditions set by demanding continuity.

\subsection{Pump-induced effective fields} \label{sec:pump-induc-BKT}

The pump-induced effective fields can be obtained by considering the bulk expression Eq.~\eqref{eq:heff_pump_bulk} restricted to the case of in-plane $\ve{n}$ and out-of-plane magnetization $m_z$, yielding
\begin{align}
	\mathcal{S}_\mathrm{eff}^\mathrm{pump,XY} = \frac{1}{2 v_\mathrm{u.c.}} &\int \du t \du^2 \ve x \ 2S \big[h_m m_z  \nonumber\\ &+ n_0 \left(h_{n,x} \cos \phi + h_{n,y} \sin \phi \right) \big].
\end{align} 
Here we note that the field $h_m$ couples linearly to $m_z$ and thus is readily incorporated as an inhomogeneity to the linear equations of motion.
Using the formalism presented in Appendix \ref{sec:linresp} and given the (classical) correlation function $C_{m_z m_z}$ we obtain the initial conditions
\begin{align} \label{eq:ic-bkt-for-m}
    m_z(\ve r, t_p^+) &\simeq 0 \\
    \partial_t m_z(\ve r, t_p^+) &\simeq \int \frac{\du^2 k}{(2 \pi)^2}  h_m(\ve k) \eu^{\iu \ve k \cdot r} \frac{\omega(k)^2 t_p}{m_\mathrm{eff}^2 S}.
\end{align}
where $\omega(k) = \sqrt{\tilde{c}^2k^2+r}$ with $\tilde{c} =  m_\mathrm{eff} \sqrt{\rho} / n_0$ and $r = m_\mathrm{eff}^2 p^2 h_p /n_0^4$. In particular, in the BKT phase, $h_p$ is irrelevant and thus $r=0$.

The corresponding initial conditions of the conjugate fields to $m_z$, $\phi$, read
\begin{align}
    \delta \phi(\ve r, t_p^+) &\simeq -\int \frac{\du^2 \ve k}{(2 \pi)^2 } \eu^{\iu \ve k \cdot \ve r} h_m(\ve k) \frac{t_p}{ n_0^2} \\
    \delta \partial_t \phi(\ve r, t_p^+) &\simeq 0  
\end{align}
Here, we have expanded in $\omega(k) t_p \ll 1$ which corresponds to the limit of ultrafast pulses.
Noting that the hydrodynamic modes in the BKT phase are described by coupled equations of $m_z$ and $\ve{\nabla} \delta\phi$ (see later discussions in Sec.~\ref{sec:pump-induc-BKT} for details), we conclude that there is a non-vanishing response only for fields $h_m(\ve k)$ which couple to non-zero wavevectors, i.e. are spatially non-uniform.

Contrary to $h_m$, the pump-induced field terms $h_{n}$ lead to nonlinear inhomogeneities in the equations of motion for $m_z$ and $\phi$.
To remedy this, one may be tempted to find a pump-induced initial condition  linear in $\phi$ using linear response theory (rather than explicitly including the inhomogeneity in the equations of motion).
However, as the correlation function $\langle \phi(r',t') \cos \phi(r,t) \rangle \equiv 0$ vanishes by the (unbroken) $\SO(2)$ symmetry both above and below $T_\mathrm{BKT}$ for $h_p =0 $, we find a vanishing linear response of the phase $\phi$ to pump-induced effective fields $h_{n,x}$ and $h_{n,y}$.

On the other hand, if $h_p > 0$ is finite (but small compared with all other energy scales in the problem), the system orders at lowest temperatures.
The low-energy excitations in this phase correspond to (gapped) fluctuations about the saddle-point field configuration of $\phi$.
We expand about one of the six degenerate minima of \eqref{eq:s-p}, $\phi_l = 2 \pi l/6$ with $l = 0, \dots, 5$, such that $\phi = \phi_l + \varepsilon \tilde{\phi}$ with $\varepsilon \ll 1$. The action for the pump-induced fields becomes (we take $l=0$ for concreteness)
\begin{equation}
	S_\mathrm{eff}^{\mathrm{pump},\mathrm{XY}} =  \frac{1}{2 v_\mathrm{u.c.}} \int \du t \du^2 \ve{x} 2 S \left[ h_m m_z + n_0 h_{n,y} \tilde{\phi}\right],
\end{equation}



This linear coupling to the pump-induced effective fields maintains the linearity of the equations of motion and can be used to determine the initial conditions for the relaxational dynamics.
We emphasize that, even though we have assumed an ordered ground state, the coupling between $h_n$ and $\tilde{\phi}$ will excite dynamics which are of XY character if the spatial profile of $h(r)$ is sufficiently non-uniform (such as a point-like laser irradiation yielding $h_n(r) \sim \delta(r)$): Such a field will in general couple to a broad range of momenta including both long and short wavelength.
At a given temperature, the anisotropy term $h_p$ becomes irrelevant at short distances and accordingly excitations with large momenta become identical to those of the XY model. In the next section, a sharper criteria for the finite frequency and momentum modes that exhibit dynamics of BKT phase is presented.

For the particular case of point-like ultrashort pulses of duration $\omega(k) t_p \ll 1$,
we find the initial conditions for $\tilde{\phi}$ its velocity (to lowest order in $t_p \omega \ll 1$)
\begin{subequations} 
\begin{align}
    \tilde{\phi}(\ve r,t_p^+) &\simeq 0 \label{eq:ic-phi} \\
    \partial_t\tilde{\phi}(\ve r, t_p^+) &\simeq \delta(\ve r)  \frac{m_\mathrm{eff}^2 S}{n_0^3} t_p h_{n_y} \label{eq:ic-dt-phi},
\end{align}
\end{subequations}
and correspondingly
\begin{subequations}
\begin{align}
    \delta m_z(\ve r,t=t_p^+) &\simeq \delta(\ve r) \frac{1}{n_0}t_p h_{n_y} \\
    \delta \partial_t m_z(\ve r,t=t_p^+) &\simeq 0,
\end{align}
\end{subequations}
where we refer the reader to Appendix \ref{sec:linresp} for technical details.

We conclude that for all phases, i.e. below and above $T_\mathrm{BKT}$ and within the ordered phase for $h_p$, pump-induced fields can excite relaxational dynamics by coupling to the out-of-plane magnetization $m_z$.
In addition, we have argued that a second (conjugate) pathway of driving spin-wave excitations with XY-model character is possible close to the transition between the vortex-paired disordered phase, and the ordered phase stabilized by a weak sixfold anisotropy $h_p$.

\subsection{Equations of motion and relaxation dynamics}

In order to model the equilibrium dynamics of the XY model (in the absence of anisotropy and pump-induced fields), we make use of the duality between the planar XY model and $2+1$-dimensional electromagnetic $\Uone$ gauge theory in the presence of charged matter as established by Ambegaokar, Halperin, Nelson and Siggia (AHNS) \cite{ambe80} as well as Coté and Griffin \cite{cote86}, with the main steps reproduced by us below.
The elecromagnetic duality allows for a unified description of the combined dynamics of spin waves and vortices in both quasi-ordered and disordered phases by making use of the appropriate constitutive relations for the dual vortex charge density and current below and above $T_\mathrm{BKT}$.

To this end, we note that the in-plane superfluid velocity $\ve u_s = \ve u_\parallel + \ve u_\perp$ can be decomposed into longitudinal and transverse components which satisfy $\nabla \times \ve u_\parallel = 0$ and $\nabla \cdot \ve u_\perp =0$.
Following Refs. \onlinecite{ambe80,cote86}, these conditions may be solved by setting $\ve u_\parallel = \nabla f$ and $\ve u_\perp = \hat{z} \times \nabla g$, where $f$ and $g$ are some smooth differentiable functions. 

The transverse component gives rise to a winding of the superfluid velocity characterized by
\begin{equation} \label{eq:def_N}
	\nabla \times \ve u_s = 2 \pi N(\ve r, t) \hat{z}
\end{equation}
which defines the vortex density $N(\ve r,t) = \sum_i n_i \delta(\ve r - \ve r_i)$ of a collection of pointlike vortices with charges $n_i$ at positions $\ve r_i$.
As vortices are stable topological objects, we have the conservation law $\partial_t N + \nabla \cdot J_v = 0$ with the vortex current $J_v = \sum_i \dot{\ve r}_i n_i \delta(\ve r- \ve r_i)$, where $n_i = \pm 1$ is the topological charge of the $i$-th vortex.
These considerations suggest the definition of the dual electric field $\ve e = \ve u_s \times \hat{z}$, such that \eqref{eq:def_N} becomes the Gauss law $\nabla \cdot \ve e = 2 \pi N_0$.
\footnote{Note that we embed the (2+1)-dimensional electromagnetic theory into the standard (3+1)-dimensional Maxwell's equation for ease of notation -- in the following we always have $\hat{z} \cdot \ve e = 0$ and $\ve b = (0,0,b)^\top$.}

The equation of motion for $m_z$ and $\phi$ from Eq.~\eqref{eq:s_xy} gives 
\begin{subequations}
    \begin{align}
    \frac{\du \phi}{\du t}&=\frac{m_{\rm eff}^2}{n_0^2}(S m_z)\label{eq:EoMxy1}\\
    \partial_t ( S m_z)&=\rho \nabla \cdot \ve u_\parallel,\label{eq:EoMxy2}
\end{align}
\end{subequations}
Eq.~\eqref{eq:EoMxy1} and~\eqref{eq:def_N} determine the time derivative of the superfluid velocity as
\begin{equation}
    \partial_t \ve u_s = \nabla \partial_t \phi - 2 \pi \hat{z}\times \ve J_v,
\end{equation}
which can be rewritten as a Ampère's law upon defining a magnetic field as $\ve b=m_\mathrm{eff}/(n_0 \sqrt{\rho}) S m_z \hat{z}$ (for details we refer the reader to Ref.~\onlinecite{cote86}).
Faraday's law follows straightforwardly from Eq.~\eqref{eq:EoMxy2}, and we further have $\nabla \cdot \ve b = 0$ since $\ve b=b(x,y)\hat{z}$.
The combined Maxwell equations for the gauge theory dual to the XY model thus read
\begin{subequations}\begin{align}
	\nabla \cdot ({\varepsilon} \ve e) &= 2 \pi \varrho_v\\
	\nabla \times \ve e &= - \frac{1}{c_0} \frac{\partial \ve b}{\partial t} \\
	\nabla \cdot \ve b &= 0 \\
	\nabla \times \ve b &= \frac{1}{c_0} \frac{\partial (\varepsilon \ve e)}{\partial t} + \frac{2 \pi}{c_0} \ve J_v,
\end{align}\end{subequations}
where $\varrho_v = N_v$ is the vortex charge density and $J_v$ the vortex current as defined above, and $c_0 = \sqrt{\rho} m_\mathrm{eff}/n_0$.
We have further introduced a (so far phenomenological) dielectric constant $\varepsilon$ to account for bound charges (see below for a detailed discussion).

It is convenient to decompose vectors in the XY-plane into their longitudinal and transverse components.
As introduced earlier, for the superfluid velocity we have $\ve u_\parallel = u_\parallel \hat{k}$ and $\ve u_\perp = u_\perp \hat{z} \times \hat{k}$, corresponding to spin-wave and vortex contributions, respectively.
The longitudinal and transverse projections of the electric field in an analogous decompostion then read $e_\parallel = u_\perp$ and $e_\perp = - u_\parallel$.
Fourier transforming, the second Maxwell equation reads (we write $k = |\ve k|$ and $\hat k = \ve k / k$)
\begin{equation}
    \iu k e_\perp = \frac{1}{c_0} \iu \omega \ve b \cdot \hat{z},
\end{equation}
making clear that the \emph{transverse} component of the electric field couples to $\ve b$, i.e. Eq.~\eqref{eq:EoMxy2}, such that the longitudinal components of $\ve e$ decouples from the equations of motion, entering only via the Gauss law $\varepsilon k e_\parallel = 2 \pi \rho_v$.
Combining the second and fourth Maxwell equation we finally arrive at an (inhomogenous) wave equation for the transverse component of the electric field, written as
\begin{equation} \label{eq:wave-eq-general}
    \left(\epsilon(\omega,\ve k)\omega^2 - c_0^2 k^2 \right) e_\perp(\omega,\ve k) = - 2 \pi \iu \omega \hat{z} \cdot (\hat k \times \ve J_v).
\end{equation}
The solution of \eqref{eq:wave-eq-general} depends on the nature of the vortex current $\ve J_v$ which can be related to the electric field via constitutive relations, taking different forms above and below $T_\mathrm{BKT}$, as we detail below.

\begin{figure}[tb]
	\includegraphics[width=\columnwidth]{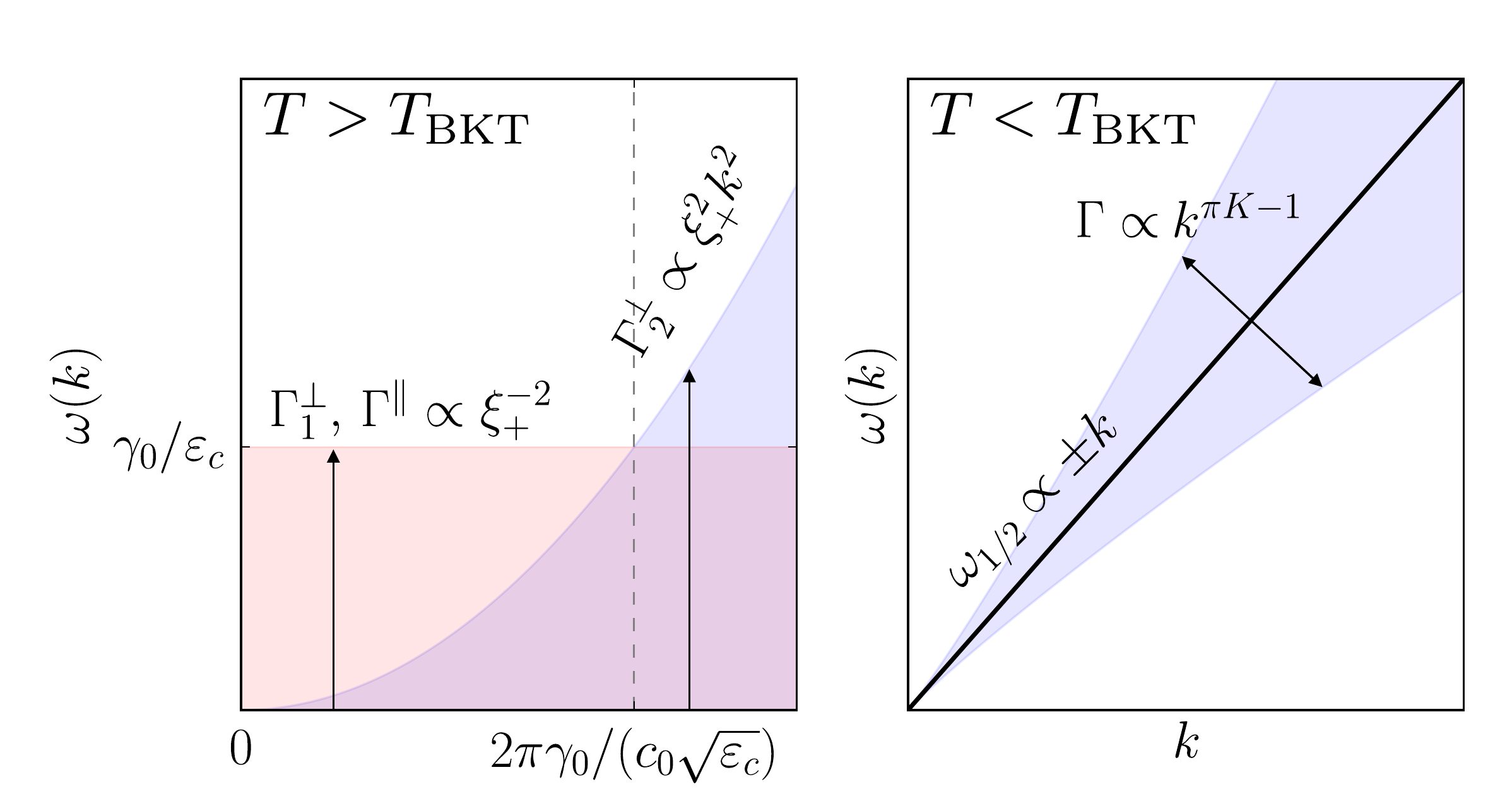}
	\caption{Qualitative illustration of the long-wavelength dynamics of the XY model. For $T> T_\mathrm{BKT}$, two decaying modes with widths $\Gamma_1^\perp$ and $\Gamma^\parallel$ as well as a diffusive mode with width $\Gamma_2^\perp$ are present, while below $T < T_\mathrm{BKT}$ the dynamics is governed by two linearly dispersing spin waves with anomalous broadening. Note that the mode $\Gamma^\parallel$ corresponds to the longitudinal electric field (vortex contributions to the superfluid velocity), whose dynamics cannot be excited through the mechanisms considered here.}
\end{figure}

\subsubsection{Gas of free vortices for $T > T_\mathrm{BKT}$}

Above $T_\mathrm{BKT}$, the proliferation of vortices is entropically favored -- within the electromagnetic theory described above, this phase is described in terms of a gas of free charges, with a renormalized dielectric function accounting for bound vortex-antivortex pairs with separations below a temperature-depdent characteristic separation scale $\xi_+(T)$ which diverges as $T \to T_\mathrm{BKT}^+$.
In order to determine the dynamics of the system, the constitutive relation for the vortex current $J_v$ must be specified.
Following AHNS \cite{ambe80}, the vortices experience a Magnus force resulting in a net transverse current as well as a stochastic force giving rise to longitudinal diffusive behaviour, such that the vortex current is written
\begin{subequations}\begin{align}
	\ve J_v (\omega, k) &= \gamma_0 \left( \epsilon_{ij} - D\frac{k_i k_l \epsilon_{l j}}{D k^2 - \iu \omega} \right) v_j \\
	&= \gamma_0 \frac{\iu \omega}{\iu \omega - D k^2} \hat{k} e_\parallel + \gamma_0 e_\perp \hat{z} \times \hat k \label{eq:j-decomposed}
\end{align}
\end{subequations}
where $\gamma_0 \sim D n_f \rho^0 / (k_\mathrm{B} T) \sim \xi_+^{-2} \sim \eu^{-\frac{b' T }{T-T_c}}$ is a phenomenological constant and $D$ is the diffusion constant.
Using this result in \eqref{eq:wave-eq-general} one arrives at
\begin{equation}
    \left(\varepsilon_c \omega^2 + 2 \pi \iu \omega \gamma_0 - c_0^2 k^2 \right) \ve e_\perp = 0.
\end{equation}
In the limit where $c_0 k \sqrt{\varepsilon_c} \ll 2 \pi \gamma_0 \sim \xi_+^{-2}$, which is applicable to all wavelengths at high temperatures and small wavelengths at successively lowered temperatures (but still $T > T_\mathrm{BKT}$), it is sufficient to approximate the dielectric function by a constant $\varepsilon(k,\omega) \simeq \varepsilon_c$, as we are interested in length scales beyond the typical free vortex separation such that the $\iu \gamma_0 \omega$ term is more important. One thus finds two modes with dispersions
\begin{equation}
    \omega_1^\perp = -  \iu \frac{2 \pi\gamma_0}{\varepsilon_c} \quad \text{and} \quad \omega_2^\perp = - \iu \frac{c_0^2 k^2}{ 2 \pi \gamma_0 }.
\end{equation}
These modes for the transverse electric field correspond (because of $e_\perp = - u_\parallel$ and $e_\parallel = u_\parallel$) to a relaxation of the longitudinal components of the superfluid velocity.
The dynamics of the longitudinal component of the electric field is obtained from the Gauss law and the continuity equation $\partial_t \varrho_v + \nabla \cdot \ve J_v = 0$, yielding $\iu \omega k e_\parallel \varepsilon = 2 \pi \ve k \cdot \ve J_v$. Using $\ve J_v$ from above we obtain the relaxation of the longitudinal component
\begin{equation}
    \omega^\parallel = - \iu \left(D k^2 + \frac{2 \pi \gamma_0}{\varepsilon_c} \right) \approx - \iu \frac{2 \pi \gamma_0}{\varepsilon_c}.
\end{equation}
Note that the modes $\omega_1^\perp$ and $\omega^\parallel$ are (to lowest order) independent of $k$ and scale down to zero as $T \to T^+_\mathrm{BKT}$.\cite{ambe80}

\subsubsection{Dielectric of vortex-antivortex pairs for $T < T_\mathrm{BKT}$}

Below $T_\mathrm{BKT}$, the vortices are bound into vortex-antivortex pairs corresponding to electric dipoles in the dual electromagnetic theory, so that the free charge density $\varrho_v = 0$ and current $\ve J_v = 0$ vanish.\cite{ambe80}
In this regime, the bound vortex pairs thus impact the dynamics through a (in general dynamic) renormalized dielectric function $\varepsilon(\omega)$, whose real and imaginary parts can be related to the static length-dependent dielectric function $\tilde{\varepsilon}(r)$ as $\Re \varepsilon(\omega) = \tilde{\varepsilon}(r=\sqrt{14D/\omega})$ and $\Im \varepsilon(\omega)=\pi/4 \left. r \frac{\du \tilde{\varepsilon}}{\du r}\right|_{r=\sqrt{14D/\omega}}$.
After the manipulations detailed in Ref.~\onlinecite{ambe80}, one arrives at the dispersion for the transverse electric field components
\begin{equation} \label{eq:disp-below-BKT}
    \omega_\pm^\perp = \pm c_3(0) k - \iu D_3 k^{\pi K -1},
\end{equation}
where $c_3(\omega) = c_0 \sqrt{\Re \varepsilon(\omega)}$ and $D_3$ is some constant.
It thus becomes clear that in the quasi-ordered phase spin waves (with a linear dispersion, first term in \eqref{eq:disp-below-BKT}) experience damping due to the background of vortex-antivortex pairs, with a continuously varying power-law momentum dependence.
Note that the absence of free charges in this regime implies that there is no dynamics of the longitudinal component of the effective electric field.

\begin{figure}[tb]
    \centering
    \includegraphics[width=\columnwidth]{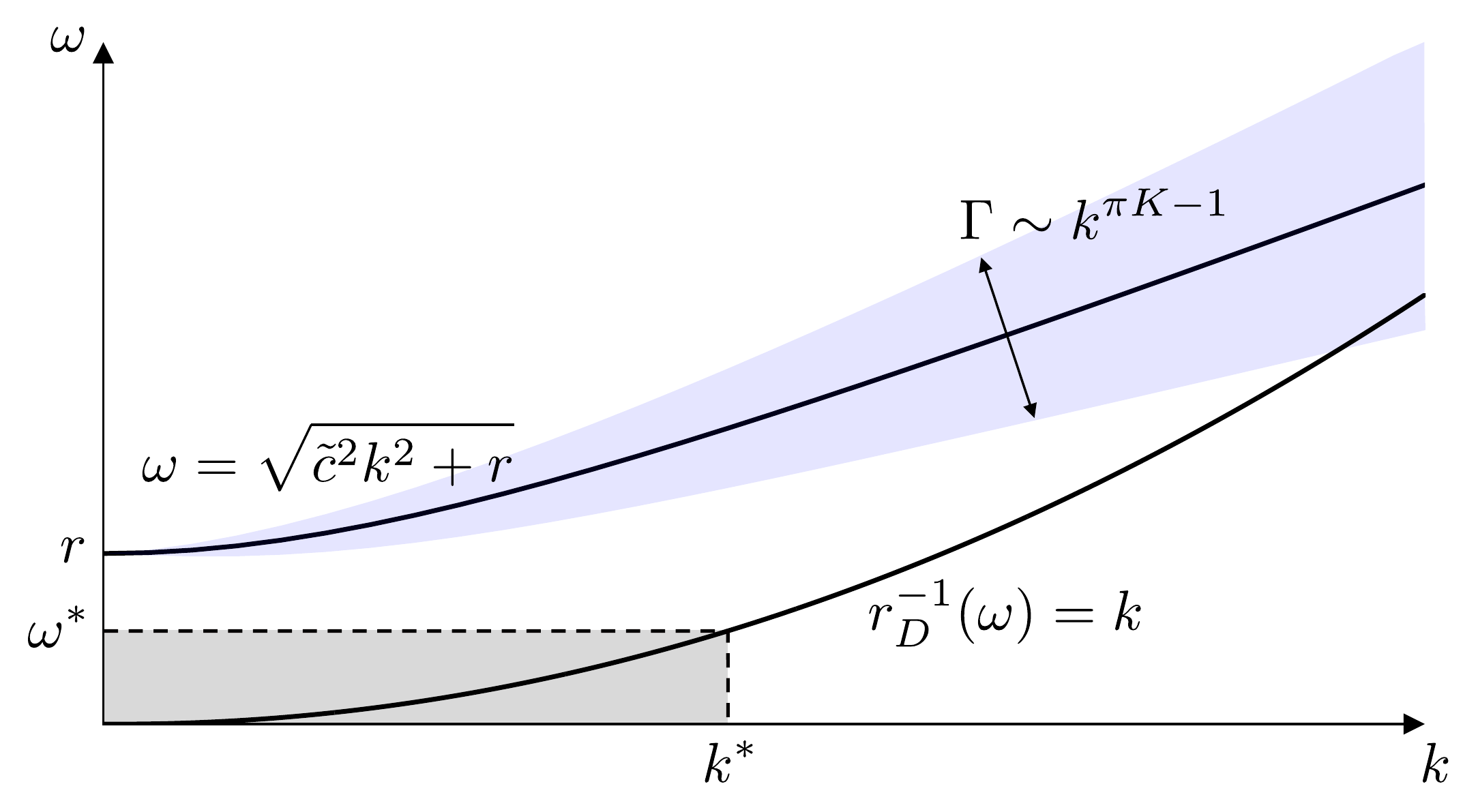}
    \caption{Illustration of BKT scaling for excited states, i.e.\ the coupled transverse electric field and magnetic field, within the ordered regime. The crossover momentum $k^\ast$ (see Fig.~\ref{fig:k-vs-T}) determines an inverse length scale which in turn sets an energy scale $\omega^\ast$ (by inverting vortex-antivortex bound pair length scale $r_D(\omega)$) above which the dielectric constant receives no further corrections from $h_p$. Thus the decay of the excited states exhibit BKT scaling (see Eq.~\eqref{eq:trans-disp-ordered}).}
    \label{fig:gapped_BKT}
\end{figure}

In passing, we argue that the dynamic renormalized dielectric function $\epsilon(\omega)$ discussed above applies to a wide range of frequency above $\omega^*\sim D k^*{}^2$ in the ordered phase close to the transition to BKT phase (see Fig.~\ref{fig:gapped_BKT}), where $k^*$ is the crossover momentum above which the static dielectric constant $\tilde \epsilon(r\sim k^{-1})$ exhibit the BKT critical power-law scaling with $k$, as illustrated in Fig.~\ref{fig:k-vs-T}.
Loosely speaking, $k^*$ can be estimated through the static RG flow of hexagonal anisotropy $h_p$~\cite{Jose1977}, around which the RG trajectory going towards the RG fixed point for $T_c$ flows to the stable fixed point for ordered phase.
Because the renormalization to $\epsilon(\omega)$ comes from the bound vortex-antivortex pairs separated within the distance $r_D(\omega)=\sqrt{14D/\omega}$, $\epsilon(\omega)$ is unchanged for $\omega>\omega^*$.
This gives the dispersion for the transverse electric field components 
\begin{equation}
    \omega^\perp_{\pm}= \pm\sqrt{{\tilde c}^2 k^2+r}-\iu D_3' ({\tilde c}^2 k^2+r)^{(\pi K-1)/2}
    \label{eq:trans-disp-ordered}
\end{equation}
for $\omega>\omega^*$.
Same as in the quasi-ordered phase, the key difference compared with magnon mode in the conventional ordered state is the damping term due to background vortex-antivortex pairs.

\subsection{Experimental consequences}

Having formulated how the pump beam's electric field gives rise to pump-induced effective fields and modelled the low-energy dynamics of the in-plane phase and out-of-plane magnetization renormalized by the presence of free vortices or bound vortex-antivortex pairs, we briefly comment on experimental implications of our studies.

For the paramagnetic (disordered) phase above $T > T_\mathrm{BKT}$, we note that no sharp quasiparticles exist, but rather only the decaying modes, two of which are momentum-independent, namely $\omega_1^\perp$ and $\omega^\parallel$. 
The remaining mode $\omega_2^\perp \sim k^2$ exhibits diffusive behaviour.
Note only the transverse modes $\omega^\perp$, which correspond to the longitudinal components of the superfluid velocity, can be excited using the microscopic mechanisms detailed earlier (this is tantamount the fact that the probe beam does not excite free vortices).
Below $T_\mathrm{BKT}$ the dynamics are given by linearly dispersing spin waves with a renormalized velocity and anomalous lifetime due to their motion in the background of vortex-antivortex pairs.

While the momentum-independent modes can in principle be probed using experimental techniques that resolve the \emph{uniform} out-of-plane magnetization $m_z$ (such as the magneto-optical Kerr effect), detection of the diffusive mode (and determining the diffusion constant empirically) as well as gapless spin waves (and their anomalous decay rates) would require probing the system at non-zero wavevectors.
One example for such an experimental protocol is given by transient grating spectroscopy which has been successfully utilized to measure spin diffusion \cite{cam96,web05} and also anomalous spin propagation \cite{web07,kora09} in semiconductors.
We further note that, as discussed in Sec.~\ref{sec:pump-induc-BKT}, highly focused pump beams (which only irradiate a small fraction of the material) couple to excitations over a broad range of momenta.
Using corresponding localized probes at different positions in the material would then also allow one to probe the dynamics of excitation with non-zero wavevectors, with details depending on the nature of the experimental setup.


\section{Conclusion}

In this work, we have established two microscopic mechanisms for driving magnetic excitations using light through the example of $\nips{}$.
While the first mechanism relies on pumping an $d$-$d$ orbital resonance {(which is rendered dipole-allowed upon considering the inversion-symmetry breaking S-induced crystal field in the crystal), splitting the $S=1$ ground-state manifold via spin-orbit coupling, we have also shown that off-resonant driving can lead to a Floquet Hamiltonian with modified exchange interactions and anisotropies for the pump duration.

In a pump-probe setup, these two mechanisms will take the system out of equilibrium and set the initial conditions for the subsequent relaxational dynamics according to the
low-energy (hydrodynamic) equations of motion. Our present work provides steps towards a more systematic understanding of the microscopic pathways how light can couple to magnetic excitations, which can be compared with experiments by studying the electric field polarization, energy and fluence dependence of the initial condition of hydrodynamic modes set by pumping. Roles of additional (intermediate) excitations such as phonons or itinerant charge carriers will be clarified in future studies. In addition, a recent publication~\cite{BelvinGedik2021} points towards pumping coherent magnons through intermediate spin-orbital entangled exciton transition. 

While the modeling of the microscopic pump mechanisms detailed above is informed by the atomic structure and nature of interactions, the relaxational dynamics (after having determined the appropriate initial conditions) depends only on the nature of the magnetic ground state and its excitations.
This implies that our framework is readily applied to structurally similar compounds which can have widely differing magnetic ground states, and be used to probe their low-energy excitations.
We have exmplified this by applying our framework to bulk $\nips{}$, finding that our study can explain the recent experimental results by Afanasiev \emph{et al.} \cite{afanasiev20}
We further suggest that pump-probe spectroscopy can be used to gain insight into monolayer $\nips{}$ which has been found to be magnetically disordered with enhanced spin fluctuations, and thus conjectured to be a magnetic realization of the XY model, potentially realizing Berezinskii-Kosterlitz-Thouless physics, which predicts strongly renormalized dynamics due to the presence of vortex-antivortex pairs and free vortices.

We hypothesize that appropriate pump-probe setups as discussed here could be of benefit in a wide range of materials for probing and controlling intrinsic coherent magnetic excitations, in particular in few-layer (and monolayer) 2D van der Waals magnets, for which established experimental techniques such as neutron scattering are inapplicable due to sample size limitations.

\acknowledgements

We thank Richard Averitt, David Hsieh, Natalia Perkins and Lucile Savary for discussions.
LB was supported by the Army Research Office MURI grant ARO W911NF-16-1-0361, Floquet engineering and metastable states. MY was supported in part by the Gordon and Betty Moore Foundation through Grant GBMF8690 to UCSB and by the National Science Foundation under Grant No.\ NSF PHY-1748958.
This project was funded in part by the European Research Council (ERC) under the European Union’s Horizon 2020 research and innovation program (Grant agreement No. 853116).

\clearpage
\onecolumngrid
\appendix

\section{Details on single-ion calculations}

\subsection{Explicit construction of multiplet wavefunctions}

Here, we explicitly construct the wavefunctions for the $A_{2g}$ and $T_{2g}$ multiplets from single-particle $d$- and $p$-orbital wavefunctions which we denote in the cubic basis by
\begin{subequations}\begin{align}
	&\ket{t_{2g}^{(i)}}_{i=1,2,3} \in \left\{\sqrt{15} yz, \sqrt{15} zx, \sqrt{15} xy\right\} \\
	&\ket{e_g^{(i)}}_{i=1,2} \in \left\{ (3 z^2 - r^2)/ \sqrt{12}, \left(x^2 -y^2\right)/2 \right\} \\
	&\ket{p^{(i)}}_{i=1,2,3} \in \left\{ \sqrt{3} x,  \sqrt{3} y,  \sqrt{3} z \right\} 
\end{align}\end{subequations}
Note that above basis states are orthonormal with respect to the inner product of spherical harmonics
$\braket{f | g} = (4 \pi)^{-1} \int_0^{2\pi} \int_{0}^\pi f^\ast(\phi,\theta) g (\phi,\theta) \sin \theta  \du \theta \du \phi$.
We then perform a unitary basis change to the $C_3$ eigenbasis with
\begin{equation}
	U = \frac{1}{\sqrt{3}} \begin{pmatrix}
		1 & \omega^2 & \omega \\
		1 & \omega & \omega^2 \\
		1 & 1 & 1
	\end{pmatrix} \oplus \frac{1}{\sqrt{2}} \begin{pmatrix}
		1 & 1 \\
		-\iu & \iu 
	\end{pmatrix} \oplus \frac{1}{\sqrt{3}} \begin{pmatrix}
		1 & \omega^2 & \omega \\
		1 & \omega & \omega^2 \\
		1 & 1 & 1 \end{pmatrix}
\end{equation}
where $\omega = \eu^{\iu 2 \pi / 3}$ such that the $C_3$ eigenstates are given by
\begin{equation} \label{eq:c3-basis}
	\left(\ket{t_1}, \ket{t_\omega},\ket{t_{\omega^2}}, \ket{e_\omega}, \ket{e_{\omega^2}}, \ket{p_1}, \dots \right) =  \left(\ket{t_{2g}^{(1)}}, \dots, \ket{p^{(3)}} \right) U.
\end{equation}
Noting that $e_g \times e_g \to {}^1A_{1g} + {}^3 A_{2g}^{(a)} + {}^1 E_g$, $e_g \times t_{2g} \to {}^1 T_{1g} + {}^3 T_{1g} + {}^1 T_{2g} + {}^3 T_{2g}$ and $t_{2g} \times t_{2g} \to {}^1 A_{1g} + {}^1 E_g + {}^3 T_{1g}$, we can then uniquely obtain $\ket{A_{2g}}$ as given in the main text and the $T_{2g}$ orbital triplet as
\begin{align} \label{eq:T2g_c3}
	\ket{T_{2g},1} &= \frac{1}{2} \left( \ket{e_\omega}_1 \ket{t_{\omega^2}}_2 +  \ket{e_{\omega^2}}_1 \ket{t_{\omega}}_2 - (1 \leftrightarrow 2) \right) \nonumber\\
	\ket{T_{2g},\omega} &= \frac{1}{2} \left( \ket{e_\omega}_1 \ket{t_{1}}_2 +  \ket{e_{\omega^2}}_1 \ket{t_{\omega^2}}_2 - (1 \leftrightarrow 2) \right) \nonumber\\
	\ket{T_{2g},\omega^2} &= \frac{1}{2} \left( \ket{e_\omega}_1 \ket{t_{\omega}}_2 +  \ket{e_{\omega^2}}_1 \ket{t_{1}}_2 - (1 \leftrightarrow 2) \right).
\end{align}
To evaluate angular momentum $L^\alpha$ and dipole $r^\alpha$ operator matrix elements (where $r^\alpha = (x,y,z)$), we find it convenient to first evaluate them in the cubic basis and reference frame, and then change to the $C_3$ eigenbasis and transform into the trigonal frame using the tranformation $W$ defined by Eq.~\eqref{eq:trafo-trigonal}, such that
\begin{equation}
	\braket{d|L^\alpha|d'} = \sum_{c,c'} V^{\alpha \beta} U^\dagger_{d c} \braket{c|L^\beta|c'} U_{c'd'},
\end{equation}
and equivalently for $\braket{d|r^\alpha|d'}$.
Here, $\ket{c}$ and $\ket{d}$ denote elements of the cubic basis and $C_3$ eigenbasis basis, respectively.

Evaluating the matrix elements of $L^\alpha = L_1^\alpha + L_2^\alpha$ in the subspace spanned by \eqref{eq:T2g_c3}, we obtain Eq.~\eqref{eq:L_projected}, where we identify $\ket{l^z = +1} \equiv \ket{T_{2g},\omega^2}$, $\ket{l^z = 0} \equiv -\ket{T_{2g},1}$ and $\ket{l^z = -1} \equiv -\ket{T_{2g},\omega}$.

\subsection{Evaluation of the effective time evolution operator} \label{app:eval-eff-U}

To evaluate the time evolution operator in \eqref{eq:ueff-1}, we represent intermediate states as a linear combination $\ket{m} = V^m_{ls} \ket{l^z}\ket{s}$ where we use the shorthand $l \equiv l^z$ and $S \equiv S^z$ with some coefficients $V^m_{ls}$ (summation convention applies).
The matrix elements then follow as
\begin{equation}
	\braket{A_2|r^\alpha|m} = \underbrace{\braket{A_2|r^\alpha|l}}_{=: M_{\alpha l}} \ket{s} V^m_{ls} = M_{\alpha l} V^m_{ls} \ket{s}.
\end{equation}
The matrix representation of the effective $3 \times 3$ Hamiltonian for the ground state $S=1$ manifold then follows as 
\begin{equation}
	U_\mathrm{eff} = \mathds{1} + \sum_m \frac{(\mathcal{E}_x,\mathcal{E}_y) M V^m \otimes (V^m)^\top M^\dagger (\mathcal{E}_x^\ast,\mathcal{E}_y^\ast)^\top}{\iu(\omega-\varepsilon_{m0})} \left[t_p - \frac{\eu^{\iu (\omega-\varepsilon_{m0})t_p}-1}{\iu (\omega- \varepsilon_{m0})}\right],
\end{equation}
where ``$\otimes$'' denotes the outer (dyadic) product.
We decompose $\mathcal{U}_\mathrm{eff} \sim  \sum_\mathcal{O} C_\mathcal{O} \mathcal{O}$ into a basis of $3\times 3$ hermitian operators $\mathcal{O}$ by taking the operator scalar product $C_\mathcal{O} = \tr\left[\mathcal{H}_\mathrm{eff} \mathcal{O}\right]/2$ where $\mathcal{O} \in \left\{\mathds{1}, S^\alpha, \{S^\alpha,S^\beta\}, (S^x)^2 - (S^y), (3 (S^z)^2 - 2 \mathds{1})/\sqrt{3} \right\}$ with $\alpha < \beta$.


\section{Details on spin-exchange calculations}
\subsection{Useful expressions}
The three-fold rotation along $\hat{z}$ in trigonal coordinate on fermion basis $\Psi$ and 3d vectors, respectively, are
\begin{align}\label{eq:C3}
W_{C_3}= \eu^{i\frac{2\pi}{3}\tau_y}\otimes_{\rm Kron} \eu^{i\frac{\pi}{3}\sigma_z} \quad \text{and} \quad 
\mathcal{R}_{C_3}=
\begin{pmatrix}
-\frac{1}{2} & \frac{\sqrt{3}}{2} & 0\\
\frac{\sqrt{3}}{2}  & -\frac{1}{2} & 0\\
0& 0 & 1 \\
\end{pmatrix}.
\end{align}
From Eqs.~\eqref{eq:effspin},~\eqref{eq:effspin1} and~\eqref{eq:C3}, the equilibrium spin exchange interaction on bonds $\ve{\delta}_{2,3}$ can be expressed explicitly as
\begin{align}\label{eq:effspin23}
\mathsf{\Gamma}^{\ve{\delta}_2}=
\begin{pmatrix}
J_l-J_l^{\prime}-\frac{1}{2}J_l^{\prime\prime} & \frac{\sqrt{3}}{2} J_l^{\prime\prime} & -\frac{1}{2} J_{l,xz} \\
\frac{\sqrt{3}}{2} J_l^{\prime\prime} & J_l-J_l^{\prime}+\frac{1}{2}J_l^{\prime\prime} & -\frac{\sqrt{3}}{2} J_{l,xz}\\
-\frac{1}{2} J_{l,xz} & -\frac{\sqrt{3}}{2} J_{l,xz} & J_l+J_l^{\prime}-J_l^{\prime\prime} \\
\end{pmatrix},
\mathsf{\Gamma}^{\ve{\delta}_3}=
\begin{pmatrix}
J_l-J_l^{\prime}-\frac{1}{2}J_l^{\prime\prime} & -\frac{\sqrt{3}}{2} J_l^{\prime\prime} & -\frac{1}{2} J_{l,xz} \\
-\frac{\sqrt{3}}{2} J_l^{\prime\prime} & J_l-J_l^{\prime}+\frac{1}{2}J_l^{\prime\prime} & \frac{\sqrt{3}}{2} J_{l,xz}\\
-\frac{1}{2} J_{l,xz} & \frac{\sqrt{3}}{2} J_{l,xz} & J_l+J_l^{\prime}-J_l^{\prime\prime} \\
\end{pmatrix},
\end{align}
where $l=1,3$ refers to NN and TNN couplings.

\section{Equation of motion for low energy spin fields}
\label{app:EoM}
Here, we derive the equation of motion (EoM) for the low energy spin fields. We will consider both the duration of the pump field and the probe period. 

We proceed with the standard non-linear sigma model (NLSM) formulation for collinear antiferromagnets, and obtain the effective Lagrangian in terms of the continuous spin fields, i.e.\ the staggered fields ($\bm{n}$), their conjugate ferromagnetic fields ($\bm{m}$). To model the effects of the pump field, we further consider the effective magnetic field $\bm{h}_n, \bm{h}_m$ that couple to $\bm{n}, \bm{m}$, respectively.

Following Ref.~\cite{SachdevBook}, the coherent state path integral for a Heisenberg spin with spin value $S$ at site $\ver$ can be obtained in the basis of the (unit) vector field $\bm{e}_\ver$, which is defined through $\vhat{S}_\ver \ket{\bm{e}_\ver}=S \bm{e}_\ver \ket{\bm{e}_\ver}$ and
$|\bm{e}_\ver|^2=1$.  For a collinear zigzag N\'eel order, it is convenient to parametrize $\bm{e}_\ver$ as
\begin{align}
\bm{e}_\ver\equiv \bm{e}_{{\bf R},\alpha}=(-1)^\alpha \eu^{i {\bf M}\cdot {\bf R}} \,\bm{n}_{\ver} \sqrt{1-(\bm{m}_{\ver})^2}+\bm{m}_{\ver},
    \label{appeq:NLSMN}
\end{align}
where ${\bf M}$ is the wave vector of the zigzag order, $\bm{n}$ is the staggered component with normalization condition $|\bm{n}|=1$, $\bm{m}$ is the uniform magnetization per site in unit of the
saturation magnetization ($=S$ semiclassically). We have defined $\ver={\bf R}+\bm{u}_\alpha$, where ${\bf R}$ is the coordinate of a unit cell, $\alpha$ labels the A, B sublattice on a honeycomb lattice. Without loss of generality, we consider the zigzag order with ${\bm M}=(0,\frac{2\pi}{\sqrt{3}})$, and $\bm{u}_A=0,\bm{u}_B=(0,-\frac{1}{\sqrt{3}})$.

The effective spin action for the spin Hamiltonian $\mathcal{H}_{\rm spin}$ is
\begin{align}
    \mathcal{Z}_s &= \int \Diff\bm{n}\Diff \bm{m} \delta(\bm{n}^2-1)\delta(\bm{n}\cdot\bm{m})\exp{(\iu\mathcal{S}_s)}\non\\
    \mathcal{S}_s & =\frac{1}{2 \uc}\int\du x \du y \du t \{2 S \bm{m}\cdot \left(\bm{n}\times \frac{\partial \bm{n}}{\partial t}\right)-m_{\rm eff}^2 S^2 \bm{m}^2 -4(D_z+D_{xy}) S^2 n_z^2-8 D_{xy} S^2 n_y^2+\rho \uc (\ve \nabla \ve{n})^2+\non\\
    &\qquad\qquad\qquad\qquad\quad 2S (\bm{h}_n \cdot \bm{n}+\bm{h}_m \cdot \bm{m})\},
    \label{appeq:spinaction}
\end{align}
where $\uc$ is the volume of the unit cell, coming from converting $\frac{1}{2}\sum_{\ver}\rightarrow \frac{1}{\uc}\int \du x \du y$, with the factor $\frac{1}{2}$ entering due to two sublattice unit cell. $m_{\rm eff}^2=4(J_1+3 J_3)$. 

To study the homogeneous order parameter dynamics, the spatial gradient on $\bm{n}, \bm{m}$ can be ignored. Below, we consider a zigzag order with the N\'eel  vector along the $x$-axis, i.e.\ $\lim_{h\rightarrow 0} \langle \bm{n}\rangle = n_0\bm{\hat{x}}$. The staggered field can be parameterized by $\bm{n}=\{n_0,n_y,n_z\}$, where $n_0=\sqrt{ 1-n_y^2-n_z^2}$ is the order parameter, and $n_{y,z}$ are
transverse fluctuations (spin waves). The first term in $\mathcal{S}_s$ comes from the Berry phase of a quantum spin operator, and in terms of $n_{y,z}$ and $m_{y,z}$, it becomes 
\begin{align}
2 S \bm{m}\cdot \left(\bm{n}\times \frac{\partial \bm{n}}{\partial t}\right) \longrightarrow -2 S n_0 \left(m_y \partial_t n_z - m_z \partial_t n_y\right),
\end{align}
which indicates that $n_y,m_z$ and $n_z, m_y$ are two sets of conjugate fields. Applying the Euler-Lagrange equation for $n_y,m_z$ and $n_z, m_y$, we arrived at the EoM for the continuous fields
\begin{align}\label{eq:EoM}
\dot{n_y}&=\chi^{-1} m_z - h_{m,z}, \quad \dot{m_z}= - 8 SD_{xy}  n_y + h_{n,y}\non\\
\dot{n_z}&=-\chi^{-1} m_y + h_{m,y}, \quad \dot{m_y}=  4S (D_z+ D_{xy})  n_z - h_{n,z}.
\end{align}
Here, $\chi$ is the uniform spin susceptibility, at the leading order in $1/S$ and ignoring the anisotropy, it is determined by $\chi_0^{-1}=S m_{\rm eff}^2/n_0 =4S(J_1+3J_3)/n_0 $. 

Eq.~\eqref{eq:EoM} can be further expressed as
\begin{align}
\partial^2_t m_z +2\gamma \partial_t m_z+ \Omega_{n_y}^2 m_z &= \kappa_{n_y} h_{m,z}+\partial_t h_{n,y}\non\\
\partial^2_t m_y + 2\gamma \partial_t m_y+\Omega_{n_z}^2 m_y &= \kappa_{n_z} h_{m,y}-\partial_t h_{n,z}
\label{eq:magnonPDE}
\end{align}
Here, we have introduced a phenomenological decay term with decay rate $\gamma$, which physically comes from coupling with the environment and satisfies the causality. The oscillation frequencies are $\Omega_{f_1}=\Omega_{n_y}=\sqrt{8S D_{xy} \chi^{-1}}=2S\sqrt{2(J_1+3J_3)D_{xy}}$, $\Omega_{f_2}=\Omega_{n_z}=\sqrt{4S (D_z+D_{xy}) \chi^{-1}}=2S\sqrt{(J_1+3J_3)(D_z+D_{xy})}$.


\section{Initial conditions in linear-response theory} \label{sec:linresp}

The initial conditions for the relaxational dynamics (according to the equilibrium equations of motions) due to the effective pump-induced fields can be computed in linear response theory.
Here we consider the pump-induced Hamiltonian of a field $h_\mathcal{O}$ coupling to the (classical) observable $\mathcal{O}$ with $\mathcal{H}^\mathrm{pump} = -(2 v_\mathrm{u.c.})^{-1} \int \mathrm{d}t \mathrm{d}^2 \ve{r} h_\mathcal{O}(t, \ve r) \mathcal{O}(t,\ve r)$.
We are interested in the response of $\mathcal{O}'$ due to the pump field $h_\mathcal{O}$, noting that $\mathcal{O}$ and $\mathcal{O}'$ are not necessarily the same. Hereafter we assume that $\langle \mathcal{O}' \rangle = 0$ in equilibrium.
The linear response of the observable $\mathcal{O}'$ is then given by
\begin{align}
    \delta \mathcal{O}'(\ve r, t) = (2 v_\mathrm{u.c.})^{-1}\int \du^2 \ve{r}' \int_{-\infty}^\infty \du t' \chi_{\mathcal{O}'\mathcal{O}}(\ve r- \ve{r}',t-t') h_\mathcal{O}(r',t')
\label{eq:ic_0}
\end{align}
where $\chi_{\mathcal{O}'\mathcal{O}}(\ve r, t)$ is the susceptibility (retarded response function) of $\mathcal{O}'$ and $\mathcal{O}$.
We assume a ``box'' temporal pump profile of the form $h_\mathcal{O}(r,t) = \bar{h}_\mathcal{O}(r) \left[\Theta(t)- \Theta(t-t_p) \right]$, where $t_p$ is the pump duration and $\bar{h}_\mathcal{O}(r)$ is the (in general spatially dependent) pump strength.
We evaluate the initial conditions shortly after the pump pulse at time $t=t_p^+$ such that $t_p^+ - t_p = 0^+$ is an infinitesimally small positive number, yielding $\delta \mathcal{O}'|_{t=t_p^+} = (2 v_\mathrm{u.c.})^{-1} \int \du^2 \ve r \int_0^{t_p} \du t'
    \bar{h}_\mathcal{O}(\ve r') \chi(\ve r- \ve r', t_p^+ -t')$
and similarly for $\partial_t \delta \mathcal{O}$. For brevity, the subscript of $\chi$ will be dropped unless there is ambiguity.
Writing the susceptibility in terms of its imaginary part $\chi''(\omega)$ in the frequency domain $\chi(t-t') = \int \du \omega' \frac{\iu}{\pi} \eu^{-\iu \omega (t-t')} \chi''(\omega) \Theta(t-t')$ (which is obtained making use of Kramers-Kronig relations and Plemelji theorem)~\cite{ColemanBook}, the temporal integration can be performed,
\begin{align} 
    \delta \mathcal{O}'|_{t=t_p^+} = (2 \uc)^{-1}\int \du^2 \ve{r}' \int \du \omega' h_\mathcal{O}(\ve r') \frac{1}{\pi \omega'} \left(1 - \eu^{-\iu t_p \omega'}\right) \chi''(\ve r - \ve r',\omega').
\label{eq:ic_1}
\end{align}
We further note that $\chi''$ is related to the correlation function $C_{\mathcal{O}'\mathcal{O}}(\ve r,t) = \langle \mathcal{O}'(\ve r,t) \mathcal{O}(0,0) \rangle$ through the fluctuation-dissipation theorem. In the classical limit it reads $2 \chi''(\omega, \ve k) = \beta \omega  C(\omega, \ve k)$, thus 
\begin{align}
       \delta \mathcal{O}'|_{t=t_p^+}(\ve r) = (2 \uc)^{-1}\int \frac{\du^2 \ve k}{(2 \pi)^2} \int \du \omega' \frac{\beta}{2 \pi} \eu^{\iu \ve k \cdot \ve r} \left(1 - \eu^{-\iu t_p \omega'}\right) \bar{h}_\mathcal{O}(\ve k) C(\omega', \ve k) 
        \label{eq:ic_2}
\end{align}
Depending the the nature of perturbing and responding observables, it is convenient to use either Eq.~\eqref{eq:ic_0},~\eqref{eq:ic_1} or~\eqref{eq:ic_2}. 

To proceed, it is convenient to relate the response function $\chi_{(\partial_t \mathcal{O}') \mathcal{O}}$ and $\chi_{\mathcal{O}' (\partial_t\mathcal{O})}$ with $\chi_{\mathcal{O}' \mathcal{O}}$ as
\begin{equation}
    \chi_{(\partial_t \mathcal{O}') \mathcal{O}}(\omega,\ve k)=-\chi_{\mathcal{O}'(\partial_t\mathcal{O})}(\omega,\ve k)=\frac{\omega}{\iu}\chi_{\mathcal{O}' \mathcal{O}}(\omega,\ve k),
    \label{eq:ptOO}
\end{equation}
where using $\chi_{(\partial_t \mathcal{O}') \mathcal{O}}$ in the linear response formulae above gives the initial condition of $\delta \partial_t \mathcal{O}'$. 
\subsection{Linear response deep in the ordered state}
To obtain the initial condition in the probe period for coherent magnon in the zigzag ordered state, we follow the discussion in Ref.~\onlinecite{seiba19}, from Eq.~\eqref{eq:magnonPDE}, the Fourier transform of spin susceptibility at $\ve k=0$ reads
\begin{equation}
    \chi_{m_z m_z}(\omega)\sim \frac{1}{\omega^2+2\iu \gamma \omega -\Omega_{n_y}^2 }
\end{equation}
In the limit of short pulse, i.e.\ $t_p \ll \omega^{-1}, \gamma^{-1} $, the effective field $\ve{h}_m$ acts as an impulse to the magnetization, giving the magnetization an initial velocity $\partial_t m $, while the effective field $\ve{h}_n$ provides initial amplitude of $\ve{m}$ for free oscillation in the probe period. We find (setting $n_0\equiv 1$)
\begin{align}
m_z (t_p^+) &= \bar{h}_{n,y} t_p ,\quad \partial_t m_z(t_p^+)= \kappa_{n_y}\bar{h}_{m,z} t_p \non\\
m_y (t_p^+) &= -\bar{h}_{n,z} t_p ,\quad \partial_t m_z(t_p^+)= \kappa_{n_z}\bar{h}_{m,y} t_p
\end{align}

\subsection{Linear response in and proximate to the BKT phase}
In and proximate to the BKT phase, it is convenient to describe the dynamics in terms of the magnetization field $m_z$ and the phase field $\phi$, which can be conveniently formulated in terms of a dual electromagnetic theory~\cite{}. Defining $\phi=\phi_{l=0}+ \tilde \phi$ (without loss of generality, we take $\phi_l=0$), the action for XY model with hexagonal anisotropy $h_p$ after integrating out $m_z$ and performing the saddle-point expansion about $\phi_{l=0}=0$ reads
\begin{equation}
	S_{\mathrm{XY}} = \frac{1}{2 v_\mathrm{u.c.}} \int \du t \du^2 \ve x \Big[ \frac{n_0^4}{m_\mathrm{eff}^2} (\partial_t \tilde{\phi})^2 -\rho n_0^2 (\nabla  \tilde{\phi})^2 - h_p p^2 \tilde{\phi}^2\Big].
\end{equation}
The classical correlation function (i.e. assuming that the mode is on shell only) is obtained as 
\begin{equation} \label{eq:cphiphi}
	C_{\tilde{\phi}\tilde{\phi}}(\ve k,\omega) =  \frac{\pi m_\mathrm{eff}^2 \uc}{\beta n_0^4 \omega(k)^2} \left[\delta(\omega-\omega(k)) + \delta(\omega+\omega(k)) \right],
\end{equation}
where $\omega(k) = \sqrt{\tilde{c}^2k^2+r}$ with $\tilde{c} =  m_\mathrm{eff} \sqrt{\rho} / n_0$ and $r = m_\mathrm{eff}^2 p^2 h_p /n_0^4$, with $\beta^{-1} = k_\mathrm{B} T$ as usual. In particular, in the BKT phase, $h_p$ is irrelevant and thus $r=0$.   

The autocorrelation function for the conjugate field $m_z$ reads
\begin{equation}
    C_{m_z m_z}(\ve k, \omega) = \frac{\pi\uc}{\beta m_\mathrm{eff}^2 S^2} \left[\delta(\omega-\omega(k)) + \delta(\omega +\omega(k)) \right].
\end{equation}

From $\mathcal{S}^{\mathrm{pump,XY}}_\mathrm{eff}$ we read off $\bar h_{m_z} = 2 S h_m$ and thus obtain
\begin{subequations}
\begin{align}
    \delta m_z(\ve r,t_p^+) &= \int \frac{\du^2 \ve k}{(2 \pi)^2} \eu^{\iu \ve k \cdot \ve r} h_m(\ve k) \frac{1}{m_\mathrm{eff}^2 S} \left(1-  \cos (\omega(k) t_p) \right) \\
    \delta \partial_t m_z(\ve r, t_p^+) &= \int \frac{\du^2 \ve k}{(2 \pi)^2 } \eu^{\iu \ve k \cdot \ve r} h_m(\ve k) \frac{1}{m_\mathrm{eff}^2 S} \omega(k) \sin (\omega(k) t_p),
\end{align}
\end{subequations}
Using the autocorrelation for $\phi$ fields from Eq.~\eqref{eq:cphiphi} and Eq.~\eqref{eq:ptOO}, the initial condition for the phase field $\delta \phi(\ve r,t_p^+)$ reads
\begin{subequations}
\begin{align}
    \delta \phi(\ve r, t_p^+) &= -\int \frac{\du^2 \ve k}{(2 \pi)^2 } \eu^{\iu \ve k \cdot \ve r} h_m(\ve k) \frac{1}{ n_0^2 \omega(k)} \sin (\omega(k) t_p)\\
    \delta \partial_t \phi(\ve r, t_p^+) &= \int \frac{\du^2 \ve k}{(2 \pi)^2 } \eu^{\iu \ve k \cdot \ve r} h_m(\ve k) \frac{1}{ n_0^2 } (1-\cos (\omega(k) t_p))
\end{align}
\end{subequations}
In the limit $t_p \omega(k)\ll 1$, the initial conditions read 
\begin{subequations}
\begin{align}
    \delta m_z (\ve k,t_p^+)&\approx 0, \quad \delta \partial_t m_z (\ve k, t_p^+)  \approx \frac{ \omega(k)^2}{m_\mathrm{eff}^2 S}h_m t_p\\
    \delta \phi (\ve k,t_p^+)&\approx -\frac{1}{n_0^2}h_m t_p, \quad \delta \partial_t \phi (\ve k, t_p^+)  \approx 0
\end{align}
\end{subequations}

In the ordered phase due to the relevant 6-fold hexagonal anisotropy perturbation $h_p$, it is argued in the main text that the pump induced field $\ve{h}_n$ may also induce hydrodynamic modes of the BKT phase at finite wave vector when $T$ is close to $T_c$. Further, with $\bar{h}_{\tilde\phi} = 2 S n_0 h_{n_y}$ and assuming a $\delta(\ve r)$-spatial profile of the pump-induced fields, we obtain using $C_{\tilde \phi \tilde\phi}$ from \eqref{eq:cphiphi}, after performing the $\omega'$-integration,
\begin{subequations}
\begin{align}
    \delta \tilde\phi(\ve r,t=t_p^+) &= \int \frac{\du^2 \ve k}{(2 \pi)^2} \eu^{\iu \ve k \cdot \ve r} h_{n_y}(\ve k)\frac{ S m_\mathrm{eff}^2}{n_0^3 \omega(k)^2} \left( 1- \cos (\omega(k) t_p ) \right) \\
    \delta \partial_t \tilde\phi(\ve r,t=t_p^+) &= \int \frac{\du^2 \ve k}{(2 \pi)^2} \eu^{\iu \ve k \cdot \ve r} h_{n_y}(\ve k)\frac{ S m_\mathrm{eff}^2}{n_0^3 \omega(k)} \sin(t_p \omega(k))
\end{align}
\end{subequations}
and
\begin{subequations}
\begin{align}
    \delta m_z(\ve r,t=t_p^+) &= \int \frac{\du^2 \ve k}{(2 \pi)^2} \eu^{\iu \ve k \cdot \ve r} h_{n_y}(\ve k)\frac{1}{n_0 \omega(k)} \sin \left(t_p \omega(k)\right) \\
    \delta \partial_t m_z(\ve r,t=t_p^+) &= \int \frac{\du^2 \ve k}{(2 \pi)^2} \eu^{\iu \ve k \cdot \ve r} h_{n_y}(\ve k)\frac{1}{n_0} \left( 1- \cos (\omega(k) t_p ) \right).
\end{align}
\end{subequations}
where $\omega(k) = \sqrt{\tilde{c}^2 k^2 + r}$, with $r$ the hexagonal anisotropy gap.
Expanding in $t_p \omega(k) \ll 1$ then leads to Eqs.~\eqref{eq:ic-phi} and \eqref{eq:ic-dt-phi}.


\bibliography{optmag-nips3-bib}

\begin{thebibliography}{37}%
\makeatletter
\providecommand \@ifxundefined [1]{%
 \@ifx{#1\undefined}
}%
\providecommand \@ifnum [1]{%
 \ifnum #1\expandafter \@firstoftwo
 \else \expandafter \@secondoftwo
 \fi
}%
\providecommand \@ifx [1]{%
 \ifx #1\expandafter \@firstoftwo
 \else \expandafter \@secondoftwo
 \fi
}%
\providecommand \natexlab [1]{#1}%
\providecommand \enquote  [1]{``#1''}%
\providecommand \bibnamefont  [1]{#1}%
\providecommand \bibfnamefont [1]{#1}%
\providecommand \citenamefont [1]{#1}%
\providecommand \href@noop [0]{\@secondoftwo}%
\providecommand \href [0]{\begingroup \@sanitize@url \@href}%
\providecommand \@href[1]{\@@startlink{#1}\@@href}%
\providecommand \@@href[1]{\endgroup#1\@@endlink}%
\providecommand \@sanitize@url [0]{\catcode `\\12\catcode `\$12\catcode
  `\&12\catcode `\#12\catcode `\^12\catcode `\_12\catcode `\%12\relax}%
\providecommand \@@startlink[1]{}%
\providecommand \@@endlink[0]{}%
\providecommand \url  [0]{\begingroup\@sanitize@url \@url }%
\providecommand \@url [1]{\endgroup\@href {#1}{\urlprefix }}%
\providecommand \urlprefix  [0]{URL }%
\providecommand \Eprint [0]{\href }%
\providecommand \doibase [0]{https://doi.org/}%
\providecommand \selectlanguage [0]{\@gobble}%
\providecommand \bibinfo  [0]{\@secondoftwo}%
\providecommand \bibfield  [0]{\@secondoftwo}%
\providecommand \translation [1]{[#1]}%
\providecommand \BibitemOpen [0]{}%
\providecommand \bibitemStop [0]{}%
\providecommand \bibitemNoStop [0]{.\EOS\space}%
\providecommand \EOS [0]{\spacefactor3000\relax}%
\providecommand \BibitemShut  [1]{\csname bibitem#1\endcsname}%
\let\auto@bib@innerbib\@empty
\bibitem [{\citenamefont {Afanasiev}\ \emph {et~al.}(2021)\citenamefont
  {Afanasiev}, \citenamefont {Hortensius}, \citenamefont {Matthiesen},
  \citenamefont {Ma{\~n}as-Valero}, \citenamefont {{\v S}i{\v s}kins},
  \citenamefont {Lee}, \citenamefont {Lesne}, \citenamefont {van~der Zant},
  \citenamefont {Steeneken}, \citenamefont {Ivanov}, \citenamefont {Coronado},\
  and\ \citenamefont {Caviglia}}]{afanasiev20}%
  \BibitemOpen
  \bibfield  {author} {\bibinfo {author} {\bibfnamefont {D.}~\bibnamefont
  {Afanasiev}}, \bibinfo {author} {\bibfnamefont {J.~R.}\ \bibnamefont
  {Hortensius}}, \bibinfo {author} {\bibfnamefont {M.}~\bibnamefont
  {Matthiesen}}, \bibinfo {author} {\bibfnamefont {S.}~\bibnamefont
  {Ma{\~n}as-Valero}}, \bibinfo {author} {\bibfnamefont {M.}~\bibnamefont {{\v
  S}i{\v s}kins}}, \bibinfo {author} {\bibfnamefont {M.}~\bibnamefont {Lee}},
  \bibinfo {author} {\bibfnamefont {E.}~\bibnamefont {Lesne}}, \bibinfo
  {author} {\bibfnamefont {H.~S.~J.}\ \bibnamefont {van~der Zant}}, \bibinfo
  {author} {\bibfnamefont {P.~G.}\ \bibnamefont {Steeneken}}, \bibinfo {author}
  {\bibfnamefont {B.~A.}\ \bibnamefont {Ivanov}}, \bibinfo {author}
  {\bibfnamefont {E.}~\bibnamefont {Coronado}},\ and\ \bibinfo {author}
  {\bibfnamefont {A.~D.}\ \bibnamefont {Caviglia}},\ }\href
  {https://doi.org/10.1126/sciadv.abf3096} {\bibfield  {journal} {\bibinfo
  {journal} {Science Advances}\ }\textbf {\bibinfo {volume} {7}},\ \bibinfo
  {pages} {eabf3096} (\bibinfo {year} {2021})}\BibitemShut {NoStop}%
\bibitem [{\citenamefont {Kirilyuk}\ \emph {et~al.}(2010)\citenamefont
  {Kirilyuk}, \citenamefont {Kimel},\ and\ \citenamefont {Rasing}}]{kiri10}%
  \BibitemOpen
  \bibfield  {author} {\bibinfo {author} {\bibfnamefont {A.}~\bibnamefont
  {Kirilyuk}}, \bibinfo {author} {\bibfnamefont {A.~V.}\ \bibnamefont
  {Kimel}},\ and\ \bibinfo {author} {\bibfnamefont {T.}~\bibnamefont
  {Rasing}},\ }\href {https://doi.org/10.1103/RevModPhys.82.2731} {\bibfield
  {journal} {\bibinfo  {journal} {Rev. Mod. Phys.}\ }\textbf {\bibinfo {volume}
  {82}},\ \bibinfo {pages} {2731} (\bibinfo {year} {2010})}\BibitemShut
  {NoStop}%
\bibitem [{\citenamefont {Basov}\ \emph {et~al.}(2017)\citenamefont {Basov},
  \citenamefont {Averitt},\ and\ \citenamefont {Hsieh}}]{baso17}%
  \BibitemOpen
  \bibfield  {author} {\bibinfo {author} {\bibfnamefont {D.~N.}\ \bibnamefont
  {Basov}}, \bibinfo {author} {\bibfnamefont {R.~D.}\ \bibnamefont {Averitt}},\
  and\ \bibinfo {author} {\bibfnamefont {D.}~\bibnamefont {Hsieh}},\ }\href
  {https://doi.org/10.1038/nmat5017} {\bibfield  {journal} {\bibinfo  {journal}
  {Nature Materials}\ }\textbf {\bibinfo {volume} {16}},\ \bibinfo {pages}
  {1077} (\bibinfo {year} {2017})}\BibitemShut {NoStop}%
\bibitem [{\citenamefont {de~la Torre}\ \emph {et~al.}(2021)\citenamefont
  {de~la Torre}, \citenamefont {Kennes}, \citenamefont {Claassen},
  \citenamefont {Gerber}, \citenamefont {McIver},\ and\ \citenamefont
  {Sentef}}]{TorreReview2021}%
  \BibitemOpen
  \bibfield  {author} {\bibinfo {author} {\bibfnamefont {A.}~\bibnamefont
  {de~la Torre}}, \bibinfo {author} {\bibfnamefont {D.~M.}\ \bibnamefont
  {Kennes}}, \bibinfo {author} {\bibfnamefont {M.}~\bibnamefont {Claassen}},
  \bibinfo {author} {\bibfnamefont {S.}~\bibnamefont {Gerber}}, \bibinfo
  {author} {\bibfnamefont {J.~W.}\ \bibnamefont {McIver}},\ and\ \bibinfo
  {author} {\bibfnamefont {M.~A.}\ \bibnamefont {Sentef}},\ }\href
  {https://doi.org/10.1103/RevModPhys.93.041002} {\bibfield  {journal}
  {\bibinfo  {journal} {Rev. Mod. Phys.}\ }\textbf {\bibinfo {volume} {93}},\
  \bibinfo {pages} {041002} (\bibinfo {year} {2021})}\BibitemShut {NoStop}%
\bibitem [{\citenamefont {Duong}\ \emph {et~al.}(2004)\citenamefont {Duong},
  \citenamefont {Satoh},\ and\ \citenamefont {Fiebig}}]{fiebig04}%
  \BibitemOpen
  \bibfield  {author} {\bibinfo {author} {\bibfnamefont {N.~P.}\ \bibnamefont
  {Duong}}, \bibinfo {author} {\bibfnamefont {T.}~\bibnamefont {Satoh}},\ and\
  \bibinfo {author} {\bibfnamefont {M.}~\bibnamefont {Fiebig}},\ }\href
  {https://doi.org/10.1103/PhysRevLett.93.117402} {\bibfield  {journal}
  {\bibinfo  {journal} {Phys. Rev. Lett.}\ }\textbf {\bibinfo {volume} {93}},\
  \bibinfo {pages} {117402} (\bibinfo {year} {2004})}\BibitemShut {NoStop}%
\bibitem [{\citenamefont {Tzschaschel}\ \emph {et~al.}(2017)\citenamefont
  {Tzschaschel}, \citenamefont {Otani}, \citenamefont {Iida}, \citenamefont
  {Shimura}, \citenamefont {Ueda}, \citenamefont {G\"unther}, \citenamefont
  {Fiebig},\ and\ \citenamefont {Satoh}}]{tzsch17}%
  \BibitemOpen
  \bibfield  {author} {\bibinfo {author} {\bibfnamefont {C.}~\bibnamefont
  {Tzschaschel}}, \bibinfo {author} {\bibfnamefont {K.}~\bibnamefont {Otani}},
  \bibinfo {author} {\bibfnamefont {R.}~\bibnamefont {Iida}}, \bibinfo {author}
  {\bibfnamefont {T.}~\bibnamefont {Shimura}}, \bibinfo {author} {\bibfnamefont
  {H.}~\bibnamefont {Ueda}}, \bibinfo {author} {\bibfnamefont {S.}~\bibnamefont
  {G\"unther}}, \bibinfo {author} {\bibfnamefont {M.}~\bibnamefont {Fiebig}},\
  and\ \bibinfo {author} {\bibfnamefont {T.}~\bibnamefont {Satoh}},\ }\href
  {https://doi.org/10.1103/PhysRevB.95.174407} {\bibfield  {journal} {\bibinfo
  {journal} {Phys. Rev. B}\ }\textbf {\bibinfo {volume} {95}},\ \bibinfo
  {pages} {174407} (\bibinfo {year} {2017})}\BibitemShut {NoStop}%
\bibitem [{\citenamefont {Seifert}\ and\ \citenamefont
  {Balents}(2019)}]{seiba19}%
  \BibitemOpen
  \bibfield  {author} {\bibinfo {author} {\bibfnamefont {U.~F.~P.}\
  \bibnamefont {Seifert}}\ and\ \bibinfo {author} {\bibfnamefont
  {L.}~\bibnamefont {Balents}},\ }\href
  {https://doi.org/10.1103/PhysRevB.100.125161} {\bibfield  {journal} {\bibinfo
   {journal} {Phys. Rev. B}\ }\textbf {\bibinfo {volume} {100}},\ \bibinfo
  {pages} {125161} (\bibinfo {year} {2019})}\BibitemShut {NoStop}%
\bibitem [{\citenamefont {Wildes}\ \emph
  {et~al.}(2015{\natexlab{a}})\citenamefont {Wildes}, \citenamefont {Simonet},
  \citenamefont {Ressouche}, \citenamefont {McIntyre}, \citenamefont {Avdeev},
  \citenamefont {Suard}, \citenamefont {Kimber}, \citenamefont
  {Lan\ifmmode~\mbox{\c{c}}\else \c{c}\fi{}on}, \citenamefont {Pepe},
  \citenamefont {Moubaraki},\ and\ \citenamefont {Hicks}}]{wild15}%
  \BibitemOpen
  \bibfield  {author} {\bibinfo {author} {\bibfnamefont {A.~R.}\ \bibnamefont
  {Wildes}}, \bibinfo {author} {\bibfnamefont {V.}~\bibnamefont {Simonet}},
  \bibinfo {author} {\bibfnamefont {E.}~\bibnamefont {Ressouche}}, \bibinfo
  {author} {\bibfnamefont {G.~J.}\ \bibnamefont {McIntyre}}, \bibinfo {author}
  {\bibfnamefont {M.}~\bibnamefont {Avdeev}}, \bibinfo {author} {\bibfnamefont
  {E.}~\bibnamefont {Suard}}, \bibinfo {author} {\bibfnamefont {S.~A.~J.}\
  \bibnamefont {Kimber}}, \bibinfo {author} {\bibfnamefont {D.}~\bibnamefont
  {Lan\ifmmode~\mbox{\c{c}}\else \c{c}\fi{}on}}, \bibinfo {author}
  {\bibfnamefont {G.}~\bibnamefont {Pepe}}, \bibinfo {author} {\bibfnamefont
  {B.}~\bibnamefont {Moubaraki}},\ and\ \bibinfo {author} {\bibfnamefont
  {T.~J.}\ \bibnamefont {Hicks}},\ }\href
  {https://doi.org/10.1103/PhysRevB.92.224408} {\bibfield  {journal} {\bibinfo
  {journal} {Phys. Rev. B}\ }\textbf {\bibinfo {volume} {92}},\ \bibinfo
  {pages} {224408} (\bibinfo {year} {2015}{\natexlab{a}})}\BibitemShut
  {NoStop}%
\bibitem [{\citenamefont {Lan\ifmmode~\mbox{\c{c}}\else \c{c}\fi{}on}\ \emph
  {et~al.}(2018)\citenamefont {Lan\ifmmode~\mbox{\c{c}}\else \c{c}\fi{}on},
  \citenamefont {Ewings}, \citenamefont {Guidi}, \citenamefont {Formisano},\
  and\ \citenamefont {Wildes}}]{lan18}%
  \BibitemOpen
  \bibfield  {author} {\bibinfo {author} {\bibfnamefont {D.}~\bibnamefont
  {Lan\ifmmode~\mbox{\c{c}}\else \c{c}\fi{}on}}, \bibinfo {author}
  {\bibfnamefont {R.~A.}\ \bibnamefont {Ewings}}, \bibinfo {author}
  {\bibfnamefont {T.}~\bibnamefont {Guidi}}, \bibinfo {author} {\bibfnamefont
  {F.}~\bibnamefont {Formisano}},\ and\ \bibinfo {author} {\bibfnamefont
  {A.~R.}\ \bibnamefont {Wildes}},\ }\href
  {https://doi.org/10.1103/PhysRevB.98.134414} {\bibfield  {journal} {\bibinfo
  {journal} {Phys. Rev. B}\ }\textbf {\bibinfo {volume} {98}},\ \bibinfo
  {pages} {134414} (\bibinfo {year} {2018})}\BibitemShut {NoStop}%
\bibitem [{\citenamefont {Gu}\ \emph {et~al.}(2019{\natexlab{a}})\citenamefont
  {Gu}, \citenamefont {Zhang}, \citenamefont {Le}, \citenamefont {Li},
  \citenamefont {Xiang},\ and\ \citenamefont {Hu}}]{gu19}%
  \BibitemOpen
  \bibfield  {author} {\bibinfo {author} {\bibfnamefont {Y.}~\bibnamefont
  {Gu}}, \bibinfo {author} {\bibfnamefont {Q.}~\bibnamefont {Zhang}}, \bibinfo
  {author} {\bibfnamefont {C.}~\bibnamefont {Le}}, \bibinfo {author}
  {\bibfnamefont {Y.}~\bibnamefont {Li}}, \bibinfo {author} {\bibfnamefont
  {T.}~\bibnamefont {Xiang}},\ and\ \bibinfo {author} {\bibfnamefont
  {J.}~\bibnamefont {Hu}},\ }\href
  {https://doi.org/10.1103/PhysRevB.100.165405} {\bibfield  {journal} {\bibinfo
   {journal} {Phys. Rev. B}\ }\textbf {\bibinfo {volume} {100}},\ \bibinfo
  {pages} {165405} (\bibinfo {year} {2019}{\natexlab{a}})}\BibitemShut
  {NoStop}%
\bibitem [{\citenamefont {Kim}\ \emph {et~al.}(2019)\citenamefont {Kim},
  \citenamefont {Lim}, \citenamefont {Lee}, \citenamefont {Lee}, \citenamefont
  {Kim}, \citenamefont {Park}, \citenamefont {Jeon}, \citenamefont {Park},
  \citenamefont {Park},\ and\ \citenamefont {Cheong}}]{kim19}%
  \BibitemOpen
  \bibfield  {author} {\bibinfo {author} {\bibfnamefont {K.}~\bibnamefont
  {Kim}}, \bibinfo {author} {\bibfnamefont {S.~Y.}\ \bibnamefont {Lim}},
  \bibinfo {author} {\bibfnamefont {J.-U.}\ \bibnamefont {Lee}}, \bibinfo
  {author} {\bibfnamefont {S.}~\bibnamefont {Lee}}, \bibinfo {author}
  {\bibfnamefont {T.~Y.}\ \bibnamefont {Kim}}, \bibinfo {author} {\bibfnamefont
  {K.}~\bibnamefont {Park}}, \bibinfo {author} {\bibfnamefont {G.~S.}\
  \bibnamefont {Jeon}}, \bibinfo {author} {\bibfnamefont {C.-H.}\ \bibnamefont
  {Park}}, \bibinfo {author} {\bibfnamefont {J.-G.}\ \bibnamefont {Park}},\
  and\ \bibinfo {author} {\bibfnamefont {H.}~\bibnamefont {Cheong}},\ }\href
  {https://doi.org/10.1038/s41467-018-08284-6} {\bibfield  {journal} {\bibinfo
  {journal} {Nature Communications}\ }\textbf {\bibinfo {volume} {10}},\
  \bibinfo {pages} {345} (\bibinfo {year} {2019})}\BibitemShut {NoStop}%
\bibitem [{\citenamefont {Jos\'e}\ \emph {et~al.}(1977)\citenamefont {Jos\'e},
  \citenamefont {Kadanoff}, \citenamefont {Kirkpatrick},\ and\ \citenamefont
  {Nelson}}]{Jose1977}%
  \BibitemOpen
  \bibfield  {author} {\bibinfo {author} {\bibfnamefont {J.~V.}\ \bibnamefont
  {Jos\'e}}, \bibinfo {author} {\bibfnamefont {L.~P.}\ \bibnamefont
  {Kadanoff}}, \bibinfo {author} {\bibfnamefont {S.}~\bibnamefont
  {Kirkpatrick}},\ and\ \bibinfo {author} {\bibfnamefont {D.~R.}\ \bibnamefont
  {Nelson}},\ }\href {https://doi.org/10.1103/PhysRevB.16.1217} {\bibfield
  {journal} {\bibinfo  {journal} {Phys. Rev. B}\ }\textbf {\bibinfo {volume}
  {16}},\ \bibinfo {pages} {1217} (\bibinfo {year} {1977})}\BibitemShut
  {NoStop}%
\bibitem [{\citenamefont {Ambegaokar}\ \emph {et~al.}(1980)\citenamefont
  {Ambegaokar}, \citenamefont {Halperin}, \citenamefont {Nelson},\ and\
  \citenamefont {Siggia}}]{ambe80}%
  \BibitemOpen
  \bibfield  {author} {\bibinfo {author} {\bibfnamefont {V.}~\bibnamefont
  {Ambegaokar}}, \bibinfo {author} {\bibfnamefont {B.~I.}\ \bibnamefont
  {Halperin}}, \bibinfo {author} {\bibfnamefont {D.~R.}\ \bibnamefont
  {Nelson}},\ and\ \bibinfo {author} {\bibfnamefont {E.~D.}\ \bibnamefont
  {Siggia}},\ }\href {https://doi.org/10.1103/PhysRevB.21.1806} {\bibfield
  {journal} {\bibinfo  {journal} {Phys. Rev. B}\ }\textbf {\bibinfo {volume}
  {21}},\ \bibinfo {pages} {1806} (\bibinfo {year} {1980})}\BibitemShut
  {NoStop}%
\bibitem [{\citenamefont {Stoudenmire}\ \emph {et~al.}(2009)\citenamefont
  {Stoudenmire}, \citenamefont {Trebst},\ and\ \citenamefont
  {Balents}}]{Stoudenmire2009}%
  \BibitemOpen
  \bibfield  {author} {\bibinfo {author} {\bibfnamefont {E.~M.}\ \bibnamefont
  {Stoudenmire}}, \bibinfo {author} {\bibfnamefont {S.}~\bibnamefont
  {Trebst}},\ and\ \bibinfo {author} {\bibfnamefont {L.}~\bibnamefont
  {Balents}},\ }\href {https://doi.org/10.1103/PhysRevB.79.214436} {\bibfield
  {journal} {\bibinfo  {journal} {Phys. Rev. B}\ }\textbf {\bibinfo {volume}
  {79}},\ \bibinfo {pages} {214436} (\bibinfo {year} {2009})}\BibitemShut
  {NoStop}%
\bibitem [{\citenamefont {van~der Ziel}\ \emph {et~al.}(1965)\citenamefont
  {van~der Ziel}, \citenamefont {Pershan},\ and\ \citenamefont
  {Malmstrom}}]{vdz65}%
  \BibitemOpen
  \bibfield  {author} {\bibinfo {author} {\bibfnamefont {J.~P.}\ \bibnamefont
  {van~der Ziel}}, \bibinfo {author} {\bibfnamefont {P.~S.}\ \bibnamefont
  {Pershan}},\ and\ \bibinfo {author} {\bibfnamefont {L.~D.}\ \bibnamefont
  {Malmstrom}},\ }\href {https://doi.org/10.1103/PhysRevLett.15.190} {\bibfield
   {journal} {\bibinfo  {journal} {Phys. Rev. Lett.}\ }\textbf {\bibinfo
  {volume} {15}},\ \bibinfo {pages} {190} (\bibinfo {year} {1965})}\BibitemShut
  {NoStop}%
\bibitem [{\citenamefont {Pershan}\ \emph {et~al.}(1966)\citenamefont
  {Pershan}, \citenamefont {van~der Ziel},\ and\ \citenamefont
  {Malmstrom}}]{pershan66}%
  \BibitemOpen
  \bibfield  {author} {\bibinfo {author} {\bibfnamefont {P.~S.}\ \bibnamefont
  {Pershan}}, \bibinfo {author} {\bibfnamefont {J.~P.}\ \bibnamefont {van~der
  Ziel}},\ and\ \bibinfo {author} {\bibfnamefont {L.~D.}\ \bibnamefont
  {Malmstrom}},\ }\href {https://doi.org/10.1103/PhysRev.143.574} {\bibfield
  {journal} {\bibinfo  {journal} {Phys. Rev.}\ }\textbf {\bibinfo {volume}
  {143}},\ \bibinfo {pages} {574} (\bibinfo {year} {1966})}\BibitemShut
  {NoStop}%
\bibitem [{\citenamefont {Kim}\ \emph {et~al.}(2018)\citenamefont {Kim},
  \citenamefont {Kim}, \citenamefont {Sandilands}, \citenamefont {Sinn},
  \citenamefont {Lee}, \citenamefont {Son}, \citenamefont {Lee}, \citenamefont
  {Choi}, \citenamefont {Kim}, \citenamefont {Park}, \citenamefont {Jeon},
  \citenamefont {Kim}, \citenamefont {Park}, \citenamefont {Park},
  \citenamefont {Moon},\ and\ \citenamefont {Noh}}]{kim18}%
  \BibitemOpen
  \bibfield  {author} {\bibinfo {author} {\bibfnamefont {S.~Y.}\ \bibnamefont
  {Kim}}, \bibinfo {author} {\bibfnamefont {T.~Y.}\ \bibnamefont {Kim}},
  \bibinfo {author} {\bibfnamefont {L.~J.}\ \bibnamefont {Sandilands}},
  \bibinfo {author} {\bibfnamefont {S.}~\bibnamefont {Sinn}}, \bibinfo {author}
  {\bibfnamefont {M.-C.}\ \bibnamefont {Lee}}, \bibinfo {author} {\bibfnamefont
  {J.}~\bibnamefont {Son}}, \bibinfo {author} {\bibfnamefont {S.}~\bibnamefont
  {Lee}}, \bibinfo {author} {\bibfnamefont {K.-Y.}\ \bibnamefont {Choi}},
  \bibinfo {author} {\bibfnamefont {W.}~\bibnamefont {Kim}}, \bibinfo {author}
  {\bibfnamefont {B.-G.}\ \bibnamefont {Park}}, \bibinfo {author}
  {\bibfnamefont {C.}~\bibnamefont {Jeon}}, \bibinfo {author} {\bibfnamefont
  {H.-D.}\ \bibnamefont {Kim}}, \bibinfo {author} {\bibfnamefont {C.-H.}\
  \bibnamefont {Park}}, \bibinfo {author} {\bibfnamefont {J.-G.}\ \bibnamefont
  {Park}}, \bibinfo {author} {\bibfnamefont {S.~J.}\ \bibnamefont {Moon}},\
  and\ \bibinfo {author} {\bibfnamefont {T.~W.}\ \bibnamefont {Noh}},\ }\href
  {https://doi.org/10.1103/PhysRevLett.120.136402} {\bibfield  {journal}
  {\bibinfo  {journal} {Phys. Rev. Lett.}\ }\textbf {\bibinfo {volume} {120}},\
  \bibinfo {pages} {136402} (\bibinfo {year} {2018})}\BibitemShut {NoStop}%
\bibitem [{\citenamefont {Kang}\ \emph {et~al.}(2020)\citenamefont {Kang},
  \citenamefont {Kim}, \citenamefont {Kim}, \citenamefont {Kim}, \citenamefont
  {Sim}, \citenamefont {Lee}, \citenamefont {Lee}, \citenamefont {Park},
  \citenamefont {Yun}, \citenamefont {Kim}, \citenamefont {Nag}, \citenamefont
  {Walters}, \citenamefont {Garcia-Fernandez}, \citenamefont {Li},
  \citenamefont {Chapon}, \citenamefont {Zhou}, \citenamefont {Son},
  \citenamefont {Kim}, \citenamefont {Cheong},\ and\ \citenamefont
  {Park}}]{kang20}%
  \BibitemOpen
  \bibfield  {author} {\bibinfo {author} {\bibfnamefont {S.}~\bibnamefont
  {Kang}}, \bibinfo {author} {\bibfnamefont {K.}~\bibnamefont {Kim}}, \bibinfo
  {author} {\bibfnamefont {B.~H.}\ \bibnamefont {Kim}}, \bibinfo {author}
  {\bibfnamefont {J.}~\bibnamefont {Kim}}, \bibinfo {author} {\bibfnamefont
  {K.~I.}\ \bibnamefont {Sim}}, \bibinfo {author} {\bibfnamefont {J.-U.}\
  \bibnamefont {Lee}}, \bibinfo {author} {\bibfnamefont {S.}~\bibnamefont
  {Lee}}, \bibinfo {author} {\bibfnamefont {K.}~\bibnamefont {Park}}, \bibinfo
  {author} {\bibfnamefont {S.}~\bibnamefont {Yun}}, \bibinfo {author}
  {\bibfnamefont {T.}~\bibnamefont {Kim}}, \bibinfo {author} {\bibfnamefont
  {A.}~\bibnamefont {Nag}}, \bibinfo {author} {\bibfnamefont {A.}~\bibnamefont
  {Walters}}, \bibinfo {author} {\bibfnamefont {M.}~\bibnamefont
  {Garcia-Fernandez}}, \bibinfo {author} {\bibfnamefont {J.}~\bibnamefont
  {Li}}, \bibinfo {author} {\bibfnamefont {L.}~\bibnamefont {Chapon}}, \bibinfo
  {author} {\bibfnamefont {K.-J.}\ \bibnamefont {Zhou}}, \bibinfo {author}
  {\bibfnamefont {Y.-W.}\ \bibnamefont {Son}}, \bibinfo {author} {\bibfnamefont
  {J.~H.}\ \bibnamefont {Kim}}, \bibinfo {author} {\bibfnamefont
  {H.}~\bibnamefont {Cheong}},\ and\ \bibinfo {author} {\bibfnamefont {J.-G.}\
  \bibnamefont {Park}},\ }\href {https://doi.org/10.1038/s41586-020-2520-5}
  {\bibfield  {journal} {\bibinfo  {journal} {Nature}\ }\textbf {\bibinfo
  {volume} {583}},\ \bibinfo {pages} {785} (\bibinfo {year}
  {2020})}\BibitemShut {NoStop}%
\bibitem [{\citenamefont {{Le Flem}}\ \emph {et~al.}(1982)\citenamefont {{Le
  Flem}}, \citenamefont {Brec}, \citenamefont {Ouvard}, \citenamefont
  {Louisy},\ and\ \citenamefont {Segransan}}]{Flem82}%
  \BibitemOpen
  \bibfield  {author} {\bibinfo {author} {\bibfnamefont {G.}~\bibnamefont {{Le
  Flem}}}, \bibinfo {author} {\bibfnamefont {R.}~\bibnamefont {Brec}}, \bibinfo
  {author} {\bibfnamefont {G.}~\bibnamefont {Ouvard}}, \bibinfo {author}
  {\bibfnamefont {A.}~\bibnamefont {Louisy}},\ and\ \bibinfo {author}
  {\bibfnamefont {P.}~\bibnamefont {Segransan}},\ }\href
  {https://doi.org/https://doi.org/10.1016/0022-3697(82)90156-1} {\bibfield
  {journal} {\bibinfo  {journal} {Journal of Physics and Chemistry of Solids}\
  }\textbf {\bibinfo {volume} {43}},\ \bibinfo {pages} {455} (\bibinfo {year}
  {1982})}\BibitemShut {NoStop}%
\bibitem [{\citenamefont {Wildes}\ \emph
  {et~al.}(2015{\natexlab{b}})\citenamefont {Wildes}, \citenamefont {Simonet},
  \citenamefont {Ressouche}, \citenamefont {McIntyre}, \citenamefont {Avdeev},
  \citenamefont {Suard}, \citenamefont {Kimber}, \citenamefont
  {Lan\ifmmode~\mbox{\c{c}}\else \c{c}\fi{}on}, \citenamefont {Pepe},
  \citenamefont {Moubaraki},\ and\ \citenamefont {Hicks}}]{Wildes15}%
  \BibitemOpen
  \bibfield  {author} {\bibinfo {author} {\bibfnamefont {A.~R.}\ \bibnamefont
  {Wildes}}, \bibinfo {author} {\bibfnamefont {V.}~\bibnamefont {Simonet}},
  \bibinfo {author} {\bibfnamefont {E.}~\bibnamefont {Ressouche}}, \bibinfo
  {author} {\bibfnamefont {G.~J.}\ \bibnamefont {McIntyre}}, \bibinfo {author}
  {\bibfnamefont {M.}~\bibnamefont {Avdeev}}, \bibinfo {author} {\bibfnamefont
  {E.}~\bibnamefont {Suard}}, \bibinfo {author} {\bibfnamefont {S.~A.~J.}\
  \bibnamefont {Kimber}}, \bibinfo {author} {\bibfnamefont {D.}~\bibnamefont
  {Lan\ifmmode~\mbox{\c{c}}\else \c{c}\fi{}on}}, \bibinfo {author}
  {\bibfnamefont {G.}~\bibnamefont {Pepe}}, \bibinfo {author} {\bibfnamefont
  {B.}~\bibnamefont {Moubaraki}},\ and\ \bibinfo {author} {\bibfnamefont
  {T.~J.}\ \bibnamefont {Hicks}},\ }\href
  {https://doi.org/10.1103/PhysRevB.92.224408} {\bibfield  {journal} {\bibinfo
  {journal} {Phys. Rev. B}\ }\textbf {\bibinfo {volume} {92}},\ \bibinfo
  {pages} {224408} (\bibinfo {year} {2015}{\natexlab{b}})}\BibitemShut
  {NoStop}%
\bibitem [{\citenamefont {{Chaudhary}}\ \emph {et~al.}(2020)\citenamefont
  {{Chaudhary}}, \citenamefont {{Ron}}, \citenamefont {{Hsieh}},\ and\
  \citenamefont {{Refael}}}]{Chaudhary2020}%
  \BibitemOpen
  \bibfield  {author} {\bibinfo {author} {\bibfnamefont {S.}~\bibnamefont
  {{Chaudhary}}}, \bibinfo {author} {\bibfnamefont {A.}~\bibnamefont {{Ron}}},
  \bibinfo {author} {\bibfnamefont {D.}~\bibnamefont {{Hsieh}}},\ and\ \bibinfo
  {author} {\bibfnamefont {G.}~\bibnamefont {{Refael}}},\ }\href@noop {}
  {\bibfield  {journal} {\bibinfo  {journal} {arXiv e-prints}\ ,\ \bibinfo
  {eid} {arXiv:2009.00813}} (\bibinfo {year} {2020})},\ \Eprint
  {https://arxiv.org/abs/2009.00813} {arXiv:2009.00813 [cond-mat.str-el]}
  \BibitemShut {NoStop}%
\bibitem [{\citenamefont {Hejazi}\ \emph {et~al.}(2019)\citenamefont {Hejazi},
  \citenamefont {Liu},\ and\ \citenamefont {Balents}}]{Hejazi2019}%
  \BibitemOpen
  \bibfield  {author} {\bibinfo {author} {\bibfnamefont {K.}~\bibnamefont
  {Hejazi}}, \bibinfo {author} {\bibfnamefont {J.}~\bibnamefont {Liu}},\ and\
  \bibinfo {author} {\bibfnamefont {L.}~\bibnamefont {Balents}},\ }\href
  {https://doi.org/10.1103/PhysRevB.99.205111} {\bibfield  {journal} {\bibinfo
  {journal} {Phys. Rev. B}\ }\textbf {\bibinfo {volume} {99}},\ \bibinfo
  {pages} {205111} (\bibinfo {year} {2019})}\BibitemShut {NoStop}%
\bibitem [{\citenamefont {Gu}\ \emph {et~al.}(2019{\natexlab{b}})\citenamefont
  {Gu}, \citenamefont {Zhang}, \citenamefont {Le}, \citenamefont {Li},
  \citenamefont {Xiang},\ and\ \citenamefont {Hu}}]{HuJ2019}%
  \BibitemOpen
  \bibfield  {author} {\bibinfo {author} {\bibfnamefont {Y.}~\bibnamefont
  {Gu}}, \bibinfo {author} {\bibfnamefont {Q.}~\bibnamefont {Zhang}}, \bibinfo
  {author} {\bibfnamefont {C.}~\bibnamefont {Le}}, \bibinfo {author}
  {\bibfnamefont {Y.}~\bibnamefont {Li}}, \bibinfo {author} {\bibfnamefont
  {T.}~\bibnamefont {Xiang}},\ and\ \bibinfo {author} {\bibfnamefont
  {J.}~\bibnamefont {Hu}},\ }\href
  {https://doi.org/10.1103/PhysRevB.100.165405} {\bibfield  {journal} {\bibinfo
   {journal} {Phys. Rev. B}\ }\textbf {\bibinfo {volume} {100}},\ \bibinfo
  {pages} {165405} (\bibinfo {year} {2019}{\natexlab{b}})}\BibitemShut
  {NoStop}%
\bibitem [{\citenamefont {Joy}\ and\ \citenamefont
  {Vasudevan}(1992)}]{Joy1992}%
  \BibitemOpen
  \bibfield  {author} {\bibinfo {author} {\bibfnamefont {P.~A.}\ \bibnamefont
  {Joy}}\ and\ \bibinfo {author} {\bibfnamefont {S.}~\bibnamefont
  {Vasudevan}},\ }\href {https://doi.org/10.1103/PhysRevB.46.5425} {\bibfield
  {journal} {\bibinfo  {journal} {Phys. Rev. B}\ }\textbf {\bibinfo {volume}
  {46}},\ \bibinfo {pages} {5425} (\bibinfo {year} {1992})}\BibitemShut
  {NoStop}%
\bibitem [{\citenamefont {Mak}\ \emph {et~al.}(2019)\citenamefont {Mak},
  \citenamefont {Shan},\ and\ \citenamefont {Ralph}}]{mak2019}%
  \BibitemOpen
  \bibfield  {author} {\bibinfo {author} {\bibfnamefont {K.~F.}\ \bibnamefont
  {Mak}}, \bibinfo {author} {\bibfnamefont {J.}~\bibnamefont {Shan}},\ and\
  \bibinfo {author} {\bibfnamefont {D.~C.}\ \bibnamefont {Ralph}},\ }\href
  {https://doi.org/10.1038/s42254-019-0110-y} {\bibfield  {journal} {\bibinfo
  {journal} {Nature Reviews Physics}\ }\textbf {\bibinfo {volume} {1}},\
  \bibinfo {pages} {646} (\bibinfo {year} {2019})}\BibitemShut {NoStop}%
\bibitem [{\citenamefont {Gibertini}\ \emph {et~al.}(2019)\citenamefont
  {Gibertini}, \citenamefont {Koperski}, \citenamefont {Morpurgo},\ and\
  \citenamefont {Novoselov}}]{gibertini20}%
  \BibitemOpen
  \bibfield  {author} {\bibinfo {author} {\bibfnamefont {M.}~\bibnamefont
  {Gibertini}}, \bibinfo {author} {\bibfnamefont {M.}~\bibnamefont {Koperski}},
  \bibinfo {author} {\bibfnamefont {A.~F.}\ \bibnamefont {Morpurgo}},\ and\
  \bibinfo {author} {\bibfnamefont {K.~S.}\ \bibnamefont {Novoselov}},\ }\href
  {https://doi.org/10.1038/s41565-019-0438-6} {\bibfield  {journal} {\bibinfo
  {journal} {Nature Nanotechnology}\ }\textbf {\bibinfo {volume} {14}},\
  \bibinfo {pages} {408} (\bibinfo {year} {2019})}\BibitemShut {NoStop}%
\bibitem [{\citenamefont {Berezinsky}(1971)}]{berezinsky70}%
  \BibitemOpen
  \bibfield  {author} {\bibinfo {author} {\bibfnamefont {V.~L.}\ \bibnamefont
  {Berezinsky}},\ }\href@noop {} {\bibfield  {journal} {\bibinfo  {journal}
  {Sov. Phys. JETP}\ }\textbf {\bibinfo {volume} {32}},\ \bibinfo {pages} {493}
  (\bibinfo {year} {1971})}\BibitemShut {NoStop}%
\bibitem [{\citenamefont {Kosterlitz}\ and\ \citenamefont
  {Thouless}(1973)}]{kosterlitz73}%
  \BibitemOpen
  \bibfield  {author} {\bibinfo {author} {\bibfnamefont {J.~M.}\ \bibnamefont
  {Kosterlitz}}\ and\ \bibinfo {author} {\bibfnamefont {D.~J.}\ \bibnamefont
  {Thouless}},\ }\href {https://doi.org/10.1088/0022-3719/6/7/010} {\bibfield
  {journal} {\bibinfo  {journal} {Journal of Physics C: Solid State Physics}\
  }\textbf {\bibinfo {volume} {6}},\ \bibinfo {pages} {1181} (\bibinfo {year}
  {1973})}\BibitemShut {NoStop}%
\bibitem [{\citenamefont {Ct\'e}\ and\ \citenamefont {Griffin}(1986)}]{cote86}%
  \BibitemOpen
  \bibfield  {author} {\bibinfo {author} {\bibfnamefont {R.}~\bibnamefont
  {Ct\'e}}\ and\ \bibinfo {author} {\bibfnamefont {A.}~\bibnamefont
  {Griffin}},\ }\href {https://doi.org/10.1103/PhysRevB.34.6240} {\bibfield
  {journal} {\bibinfo  {journal} {Phys. Rev. B}\ }\textbf {\bibinfo {volume}
  {34}},\ \bibinfo {pages} {6240} (\bibinfo {year} {1986})}\BibitemShut
  {NoStop}%
\bibitem [{Note1()}]{Note1}%
  \BibitemOpen
  \bibinfo {note} {Note that we embed the (2+1)-dimensional electromagnetic
  theory into the standard (3+1)-dimensional Maxwell's equation for ease of
  notation -- in the following we always have $\protect \hat {z} \cdot
  {\protect \bm {e}} = 0$ and ${\protect \bm {b}} = (0,0,b)^\top
  $.}\BibitemShut {Stop}%
\bibitem [{\citenamefont {Cameron}\ \emph {et~al.}(1996)\citenamefont
  {Cameron}, \citenamefont {Riblet},\ and\ \citenamefont {Miller}}]{cam96}%
  \BibitemOpen
  \bibfield  {author} {\bibinfo {author} {\bibfnamefont {A.~R.}\ \bibnamefont
  {Cameron}}, \bibinfo {author} {\bibfnamefont {P.}~\bibnamefont {Riblet}},\
  and\ \bibinfo {author} {\bibfnamefont {A.}~\bibnamefont {Miller}},\ }\href
  {https://doi.org/10.1103/PhysRevLett.76.4793} {\bibfield  {journal} {\bibinfo
   {journal} {Phys. Rev. Lett.}\ }\textbf {\bibinfo {volume} {76}},\ \bibinfo
  {pages} {4793} (\bibinfo {year} {1996})}\BibitemShut {NoStop}%
\bibitem [{\citenamefont {Weber}\ \emph {et~al.}(2005)\citenamefont {Weber},
  \citenamefont {Gedik}, \citenamefont {Moore}, \citenamefont {Orenstein},
  \citenamefont {Stephens},\ and\ \citenamefont {Awschalom}}]{web05}%
  \BibitemOpen
  \bibfield  {author} {\bibinfo {author} {\bibfnamefont {C.~P.}\ \bibnamefont
  {Weber}}, \bibinfo {author} {\bibfnamefont {N.}~\bibnamefont {Gedik}},
  \bibinfo {author} {\bibfnamefont {J.~E.}\ \bibnamefont {Moore}}, \bibinfo
  {author} {\bibfnamefont {J.}~\bibnamefont {Orenstein}}, \bibinfo {author}
  {\bibfnamefont {J.}~\bibnamefont {Stephens}},\ and\ \bibinfo {author}
  {\bibfnamefont {D.~D.}\ \bibnamefont {Awschalom}},\ }\href@noop {} {\bibfield
   {journal} {\bibinfo  {journal} {Nature}\ }\textbf {\bibinfo {volume}
  {437}},\ \bibinfo {pages} {1330} (\bibinfo {year} {2005})}\BibitemShut
  {NoStop}%
\bibitem [{\citenamefont {Weber}\ \emph {et~al.}(2007)\citenamefont {Weber},
  \citenamefont {Orenstein}, \citenamefont {Bernevig}, \citenamefont {Zhang},
  \citenamefont {Stephens},\ and\ \citenamefont {Awschalom}}]{web07}%
  \BibitemOpen
  \bibfield  {author} {\bibinfo {author} {\bibfnamefont {C.~P.}\ \bibnamefont
  {Weber}}, \bibinfo {author} {\bibfnamefont {J.}~\bibnamefont {Orenstein}},
  \bibinfo {author} {\bibfnamefont {B.~A.}\ \bibnamefont {Bernevig}}, \bibinfo
  {author} {\bibfnamefont {S.-C.}\ \bibnamefont {Zhang}}, \bibinfo {author}
  {\bibfnamefont {J.}~\bibnamefont {Stephens}},\ and\ \bibinfo {author}
  {\bibfnamefont {D.~D.}\ \bibnamefont {Awschalom}},\ }\href
  {https://doi.org/10.1103/PhysRevLett.98.076604} {\bibfield  {journal}
  {\bibinfo  {journal} {Phys. Rev. Lett.}\ }\textbf {\bibinfo {volume} {98}},\
  \bibinfo {pages} {076604} (\bibinfo {year} {2007})}\BibitemShut {NoStop}%
\bibitem [{\citenamefont {Koralek}\ \emph {et~al.}(2009)\citenamefont
  {Koralek}, \citenamefont {Weber}, \citenamefont {Orenstein}, \citenamefont
  {Bernevig}, \citenamefont {Zhang}, \citenamefont {Mack},\ and\ \citenamefont
  {Awschalom}}]{kora09}%
  \BibitemOpen
  \bibfield  {author} {\bibinfo {author} {\bibfnamefont {J.~D.}\ \bibnamefont
  {Koralek}}, \bibinfo {author} {\bibfnamefont {C.~P.}\ \bibnamefont {Weber}},
  \bibinfo {author} {\bibfnamefont {J.}~\bibnamefont {Orenstein}}, \bibinfo
  {author} {\bibfnamefont {B.~A.}\ \bibnamefont {Bernevig}}, \bibinfo {author}
  {\bibfnamefont {S.-C.}\ \bibnamefont {Zhang}}, \bibinfo {author}
  {\bibfnamefont {S.}~\bibnamefont {Mack}},\ and\ \bibinfo {author}
  {\bibfnamefont {D.~D.}\ \bibnamefont {Awschalom}},\ }\href
  {https://doi.org/10.1038/nature07871} {\bibfield  {journal} {\bibinfo
  {journal} {Nature}\ }\textbf {\bibinfo {volume} {458}},\ \bibinfo {pages}
  {610} (\bibinfo {year} {2009})}\BibitemShut {NoStop}%
\bibitem [{\citenamefont {{Belvin}}\ \emph {et~al.}(2021)\citenamefont
  {{Belvin}}, \citenamefont {{Baldini}}, \citenamefont {{Ozge Ozel}},
  \citenamefont {{Mao}}, \citenamefont {{Po}}, \citenamefont {{Allington}},
  \citenamefont {{Son}}, \citenamefont {{Kim}}, \citenamefont {{Kim}},
  \citenamefont {{Hwang}}, \citenamefont {{Kim}}, \citenamefont {{Park}},
  \citenamefont {{Senthil}},\ and\ \citenamefont {{Gedik}}}]{BelvinGedik2021}%
  \BibitemOpen
  \bibfield  {author} {\bibinfo {author} {\bibfnamefont {C.~A.}\ \bibnamefont
  {{Belvin}}}, \bibinfo {author} {\bibfnamefont {E.}~\bibnamefont {{Baldini}}},
  \bibinfo {author} {\bibfnamefont {I.}~\bibnamefont {{Ozge Ozel}}}, \bibinfo
  {author} {\bibfnamefont {D.}~\bibnamefont {{Mao}}}, \bibinfo {author}
  {\bibfnamefont {H.~C.}\ \bibnamefont {{Po}}}, \bibinfo {author}
  {\bibfnamefont {C.~J.}\ \bibnamefont {{Allington}}}, \bibinfo {author}
  {\bibfnamefont {S.}~\bibnamefont {{Son}}}, \bibinfo {author} {\bibfnamefont
  {B.~H.}\ \bibnamefont {{Kim}}}, \bibinfo {author} {\bibfnamefont
  {J.}~\bibnamefont {{Kim}}}, \bibinfo {author} {\bibfnamefont
  {I.}~\bibnamefont {{Hwang}}}, \bibinfo {author} {\bibfnamefont {J.~H.}\
  \bibnamefont {{Kim}}}, \bibinfo {author} {\bibfnamefont {J.-G.}\ \bibnamefont
  {{Park}}}, \bibinfo {author} {\bibfnamefont {T.}~\bibnamefont {{Senthil}}},\
  and\ \bibinfo {author} {\bibfnamefont {N.}~\bibnamefont {{Gedik}}},\ }\href
  {https://doi.org/10.1038/s41467-021-25164-8} {\bibfield  {journal} {\bibinfo
  {journal} {Nature Communications}\ }\textbf {\bibinfo {volume} {12}},\
  \bibinfo {pages} {4837} (\bibinfo {year} {2021})}\BibitemShut {NoStop}%
\bibitem [{\citenamefont {Sachdev}(2009)}]{SachdevBook}%
  \BibitemOpen
  \bibfield  {author} {\bibinfo {author} {\bibfnamefont {S.}~\bibnamefont
  {Sachdev}},\ }\href {https://doi.org/10.1017/cbo9780511973765} {\emph
  {\bibinfo {title} {Quantum Phase Transitions}}}\ (\bibinfo  {publisher}
  {Cambridge University Press},\ \bibinfo {year} {2009})\BibitemShut {NoStop}%
\bibitem [{\citenamefont {Coleman}(2015)}]{ColemanBook}%
  \BibitemOpen
  \bibfield  {author} {\bibinfo {author} {\bibfnamefont {P.}~\bibnamefont
  {Coleman}},\ }\href {https://doi.org/10.1017/CBO9781139020916} {\emph
  {\bibinfo {title} {Introduction to Many-Body Physics}}}\ (\bibinfo
  {publisher} {Cambridge University Press},\ \bibinfo {year}
  {2015})\BibitemShut {NoStop}%
\end{thebibliography}%

\bibliographystyle{apsrev4-2}

\end{document}